\documentclass[longabstract,preprint]{aastex}
\slugcomment{Astrophysical Journal}

\shorttitle{PLZ relation of RR Lyrae stars in the LMC}
\shortauthors{Muraveva et al.}

\usepackage{natbib}
\usepackage{amstext}
\usepackage{mathdesign}
\usepackage{amsmath}
\usepackage{amssymb,mathrsfs}
\usepackage[varg]{txfonts}
\usepackage{bm}
\usepackage[T1]{fontenc}
\newcommand{\setofall}[3]{\{{#1}\}_{{#2}}^{{#3}}}
\newcommand{\transpose}[1]{{#1}^{\scriptscriptstyle \top}}
\newcommand{\inverse}[1]{{#1}^{-1}}
\newcommand{\mmatrix}[1]{\boldsymbol{#1}}

\newcommand{\like}{\mathscr{L}}
\newcommand{\allz}{\setofall{z_i}{i=1}{N}}
\newcommand{\allxy}{\setofall{x_i,y_i}{i=1}{N}}
\newcommand{\allerrors}{\setofall{\sigma_{x_i},\sigma_{y_i},\sigma_{z_i}}{i=1}{N}}
\newcommand{\mS}{\mmatrix{S}}
\newcommand{\mZ}{\mmatrix{Z}}

\newcommand{\vhat}{\mmatrix{\hat{v}}}

\begin{document}

\title {New near-infrared period-luminosity-metallicity relations for RR Lyrae stars and the outlook for Gaia\footnote{Based on observations made with VISTA at ESO under programme ID 179.B-2003.}}

\author{T. Muraveva}
\affil{INAF-Osservatorio Astronomico di Bologna, via Ranzani 1, 40127, Bologna, Italy}
\email{tatiana.muraveva@oabo.inaf.it}

\author{M. Palmer}
\affil{Dept. d'Astronomia i Meteorologia, Institut de Ci\`{e}ncies del Cosmos, Universitat de Barcelona (IEEC-UB), Mart\'{i} Franqu\`{e}s 1, E08028 Barcelona, Spain}

\author{G. Clementini}
\affil{INAF-Osservatorio Astronomico di Bologna, via Ranzani 1, 40127, Bologna, Italy}

\author{X. Luri}
\affil{Dept. d'Astronomia i Meteorologia, Institut de Ci\`{e}ncies del Cosmos, Universitat de Barcelona (IEEC-UB), Mart\'{i} Franqu\`{e}s 1, E08028 Barcelona, Spain}

\author{M.-R.L. Cioni\altaffilmark{2,3}}
\affil{Universit\"{a}t Potsdam, Institut f\"{u}r Physik und Astronomie, Karl-Liebknecht-Str. 24/25, 14476 Potsdam, Germany}
\altaffiltext{2}{Leibniz-Institut f\"{u}r Astrophysik Potsdam, An der Sternwarte 16, 14482 Potsdam, Germany}
\altaffiltext{3}{University of Hertfordshire, Physics Astronomy and Mathematics, College Lane, Hatfeild AL10 9AB, 
United Kingdom}

\author{M. I. Moretti\altaffilmark{4}}
\affil{INAF-Osservatorio Astronomico di Bologna, via Ranzani 1, 40127, Bologna, Italy}
\altaffiltext{4}{Scuola Normale Superiore, Piazza dei Cavalieri 7, I-56126
Pisa, Italy}

\author{M. Marconi}
\affil{INAF-Osservatorio Astronomico di Capodimonte, Salita Moiariello 16, I-80131, Napoli, Italy}

\author{V. Ripepi}
\affil{INAF-Osservatorio Astronomico di Capodimonte, Salita Moiariello 16, I-80131, Napoli, Italy}

\author{S. Rubele}
\affil{INAF-Osservatorio Astronomico di Padova, Vicolo dell'Osservatorio 5, 35122, Padova, Italy}

\newpage
\begin{abstract}
We present results of the analysis of 70 RR Lyrae stars located in the bar of the Large Magellanic Cloud (LMC). Combining spectroscopically determined metallicity of these stars from the literature 
 with precise periods from the OGLE~III catalogue and multi-epoch $K_{\rm s}$ photometry from the {\it VISTA survey of the Magellanic Clouds system (VMC)}, we derive a new near-infrared period-luminosity-metallicity (${\rm PL_{K_{\rm s}}Z}$) relation for RR Lyrae variables. In order to fit the relation we use a fitting method developed specifically for this study. The zero-point of the relation  is estimated in two different ways: by assuming the value of the distance to the LMC 
 and by using {\it Hubble Space Telescope (HST)} parallaxes of five RR Lyrae stars in the Milky Way (MW). The difference in distance moduli derived by applying these two approaches is~$\sim0.2$ mag. To investigate this point further we derive the ${\rm PL_{K_{\rm s}}Z}$ 
 relation based on 23 MW RR Lyrae stars which had been analysed in Baade-Wesselink studies. We compared the derived ${\rm PL_{K_{\rm s}}Z}$ relations for RR Lyrae stars in the MW and LMC. Slopes and zero-points are different, but still consistent within the errors. The shallow slope of the metallicity term is confirmed by both  LMC and MW variables.
 
The astrometric space mission Gaia is expected to provide a huge contribution to the determination of the RR Lyrae ${\rm PL_{K_{\rm s}}Z}$ relation, however, calculating an absolute magnitude from the trigonometric parallax of each star and fitting a ${\rm PL_{K_{\rm s}}Z}$ relation directly to period and absolute magnitude leads to biased results. We present a tool 
to achieve an unbiased solution by modelling the data and inferring the slope and zero-point of the relation via statistical methods. 
\end{abstract}


\section{Introduction}

RR Lyrae stars are radially pulsating variables connected with low-mass helium-burning stars on the horizontal branch (HB) of the colour-magnitude diagram (CMD). These objects are Population II stars, which are abundant in  globular clusters and in the halos of galaxies. RR Lyrae stars are a perfect tool for studying  the age, formation and structure of their parent stellar system. Moreover, they are widely used for the determination of distances in the Milky Way (MW) and to Local Group galaxies.

RR Lyrae stars are primary distance indicators because of the existence of a narrow luminosity-metallicity (${\rm M_{V}- [Fe/H]}$) relation in the visual band and of period-luminosity-metallicity (${\rm PLZ}$) relations in the infrared passbands.
The near-infrared ${\rm {\rm PL_{K_{\rm s}}Z}}$ relation of RR Lyrae stars was originally discovered by \citet{Long1986}, and later was the subject of study by many different authors (e.g.,  \citealt{Bono2003},  \citealt{Cat2004},  \citealt{DelP2006}, \citealt{Sol2006}, \citealt{Sol2008}, \citealt{Bor2009}, \citealt{Cop2011}, \citealt{Ripepi2012a}).
The near-infrared ${\rm PL_{K_{\rm s}}Z}$ relation has many advantages in comparison with the visual ${\rm M_{V}-{\rm [Fe/H]}}$ relation. First of all, the luminosity of RR Lyrae stars in the $K_{\rm s}$ passband is less dependent on metallicity and interstellar extinction ($A_{K_{\rm s}} \sim 0.1A_V$).  Furthermore,  light curves of RR Lyrae stars in the $K_{\rm s}$ band have smaller amplitudes and are more symmetrical than optical light curves,  making the determination of the mean $K_{\rm s}$ magnitudes easier and more precise. 

In order to calibrate the ${\rm PL_{K_{\rm s}}Z}$ relation a large sample of RR Lyrae stars is required, spanning a wide range of metallicities, for which accurate  mean $K_{\rm s}$  and [Fe/H] measurements are available. We have selected 70 RR Lyrae variables in the Large Magellanic Cloud (LMC) with spectroscopically determined metallicities in the range of  $-2.06 < {\rm [Fe/H]}  < -$0.63  dex \citep{Grat2004}. All of them have counterparts in the OGLE~III catalogue \citep{Sosz2009}, therefore very precise periods are available. In order to increase the accuracy of the determination of mean $K_{\rm s}$ magnitudes, multi-epoch photometry is needed. For this reason we are using data from the near-infrared {\it VISTA survey of the Magellanic Clouds System (VMC}, \citealt{Cioni2011}), which is performing $K_{\rm s}$-band  observations of the whole Magellanic System in 12 (or  more) epochs, while in many previous studies only single-epoch photometry from the Two Micron All-Sky  Survey (2MASS, \citealt{Cutri2003}) was used. To fit the ${\rm PL_{K_{\rm s}}Z}$ relation we apply a fitting approach developed for the current study. This method takes into account errors in two dimensions, the intrinsic dispersion of the data and the possibility of inaccuracy in the formal error estimates.

One main issue in the determination of distances with the RR Lyrae  ${\rm PL_{K_{\rm s}}Z}$ relation is the calibration of the zero-point. Trigonometric parallaxes remain the only direct method of determining distances to astronomical sources, free of any assumptions (such as, for instance,  the distance to the LMC, etc.) and hence calibrating the  ${\rm PL_{K_{\rm s}}Z}$ zero-point.  However,  reasonably well estimated parallaxes exist, so far, only for five RR Lyrae variables in the MW observed by \cite{Ben2011} with the  {\it Hubble Space Telescope} Fine Guidance Sensor ({\it HST}/FGS). In this study we use both a global estimate of the LMC distance and the {\it HST} parallaxes in order to calibrate the zero-point of our  ${\rm PL_{K_{\rm s}}Z}$ relation based on LMC RR Lyrae stars.
Furthermore, to check whether the RR Lyrae ${\rm PL_{K_{\rm s}}Z}$ relation is universal and could thus be applied to measure distances in the MW and to other galaxies, we analyse a sample of 23 MW RR Lyrae stars, for which absolute magnitudes in the $K$ and $V$ passbands are available from the Baade-Wesselink studies (e.g., \citealt{Fern1998b}, and references therein). Based on 
these absolute magnitudes and applying our fitting approach we fit the RR Lyrae ${\rm PL_{K_{\rm s}}Z}$ relation. 
Then we compare the  ${\rm PL_{K_{\rm s}}Z}$ relations derived for RR Lyrae stars in the MW and in the LMC. 

Gaia,  the  European Space Agency (ESA) cornerstone  mission launched in December 2013, is  expected to provide a great contribution to the determination of the RR Lyrae ${\rm PL_{K_{\rm s}}Z}$  relation and to the definition of its zero-point in particular. 
The satellite is designed to produce the most precise three-dimensional (3D) map of the MW to date \citep{gaia} by measuring parallaxes of over one billion stars during its five-year mission, among which are thousands of RR Lyrae variables. In the current study we present a method which avoids the problems of the non-linear transformation of trigonometric parallaxes (and negative parallaxes) to absolute magnitudes, and apply this method to fit the ${\rm PL_{K_{\rm s}}Z}$  relation of the 23 MW RR Lyrae stars,  based on simulated Gaia parallaxes. 

In Section 2 we provide information about the 70 RR Lyrae stars in the LMC that form the basis of the present study. In Section 3 we present our method and results of fitting the RR Lyrae ${\rm PL_{K_{\rm s}}Z}$ relation in the LMC and in the MW. In Section 4 we present the method to fit the ${\rm PL_{K_{\rm s}}Z}$ relation with simulated Gaia parallaxes and apply this method to  the 23 MW RR Lyrae stars analysed in Section 3. Section 5 provides a summary of the results. In the Appendix sections we present a  detailed description of the fitting method which was developed for this study (Appendix A) and a compilation of metal abundances for the MW RR Lyrae stars (Appendix B).

\section{Data\label{dat}}
Optical photometry for the LMC RR Lyrae stars discussed in this paper was obtained by \citet{Clem2003} and \citet{Fab2005} 
using the Danish 1.54 meter telescope in La Silla, Chile. Two different sky positions, hereafter called fields A and B were observed.  Both are located close to the bar of the LMC (\citealt{Clem2003}, \citealt{Fab2005}). As a result, accurate {\it B}, {\it V} and {\it I} light curves tied to the Johnson-Cousins standard system and pulsation characteristics (period, epoch of maximum light,  amplitudes and mean magnitudes) for 125 RR Lyrae stars were obtained \citep{Fab2005}. 
Low-resolution spectra for 98 of these variables were 
collected by \citet{Grat2004} using the FOcal Reducer/low dispersion Spectrograph (FORS1) instrument mounted at the ESO VLT. They were used to derive metal abundances                      
for individual stars by comparing the strength of the Ca II K line with that of the H lines \citep{Prest1959}. For the calibration of the method, four clusters with metallicity in the range [$-$2.06; $-$1.26] dex were used. According  to  \citet{Grat2004}, 
the obtained metallicities are tied to a scale, which is, on average, 0.06 dex more metal-rich than the \citet{ZW1984} metallicity scale.

We cross-matched the sample of 98 RR Lyrae variables with known metallicities against the catalogue of RR Lyrae stars observed by the OGLE~III survey \citep{Sosz2009}. The OGLE~III catalogue contains information about the 
position, photometric and pulsation properties of 24906 RR Lyrae stars in the LMC. 
We found that, respectively, 94, 2 and 2 objects are cross-identified
with sources in the OGLE~III catalogue within a pairing radius of $1\arcsec$, $3\arcsec$ and
$7\arcsec$. The 2 stars with a counterpart at more than  $5\arcsec$ are OGLE-LMC-RRLYR-10345 and
OGLE-LMC-RRLYR-10509; for these two objects we checked both the OGLE~III
finding charts and \citet{Grat2004} Figure~5 (field B1) in order
to understand if they are affected by any problem.
Star OGLE-LMC-RRLYR-10345 is an isolated slightly elongated star without any
clear blending problem, while star OGLE-LMC-RRLYR-10509 is very close to
another source possibly making more difficult to locate accurately the star center. Considering that \citet{Grat2004} and OGLE~III periods for these 2 stars agree within 0.5\%, we kept these stars in our sample.

 We compared the periods of the 98 RR Lyrae stars provided by \citet{Fab2005} and those in the OGLE~III catalogue \citep{Sosz2009}. For 96 objects the periods agree within $\sim2$\%, while for two objects periods differ significantly. For  star A6332 the difference is of $\sim25$\% and for star A5148 it is  of $\sim37$\% (star identifications are from \citealt{Fab2005}). Moreover, star A5148 has been classified as a first-overtone RR Lyrae star (RRc) in the OGLE~III catalogue, and as a fundamental-mode RR Lyrae (RRab) by \citet{Fab2005}.   Since accurately estimated periods and classifications play a key role in the current study, we discarded these two objects from the following analysis.

Seven objects (B2811, B4008, B3625, B2517, A2623, A2119, A10360) from the sample are classified as RRc by 
\citet{Fab2005} 
and as second-overtone RR Lyrae star (RRe) in the OGLE~III catalogue. We removed them from our analysis because of the uncertain classification. Furthermore, since one of the main purposes of the current research is to study the ${\rm PL_{K_{\rm s}}Z}$ relation of RR Lyrae stars of ab- and c-types we discarded seven objects, which were classified as double-mode RR Lyrae stars (RRd) by \citet{Fab2005}: A7137, A8654, A3155, A4420, B7467, B6470 and B3347.  This left us with a final sample of 61 RRab and 21 RRc stars, which all have a counterpart in the OGLE~III catalogue. The period search for the RR Lyrae stars in the OGLE~III catalogue 
was performed using 
an algorithm based on the Fourier analysis of the light curves \citep{Sosz2009}. The uncertainties in the OGLE~III periods for the 82 RR Lyrae stars in our sample are declared to be less than $5\times 10^{-6}$ days. Therefore we used the periods provided by the OGLE~III catalogue in order to fit the ${\rm PL_{K_{\rm s}}Z}$ relation for our sample, and did not consider errors in the periods since they are negligible in comparison to the other uncertainties. 
 
In order to derive mean $K_{\rm s}$ magnitudes for the  RR Lyrae stars in our sample we used data from the VMC survey \citep{Cioni2011}. Started in 2009, the VMC survey covers a total area of 116 $\deg^2$ in the LMC with 68 contiguous tiles. The survey is obtaining {\it YJ$K_{\rm s}$} photometry. The  $K_{\rm s}$-band photometry is taken in time-series mode over 12 (or more) separate epochs and each single epoch reaches a limiting $K_{\rm s}$ magnitude $\sim19.3$ mag with a S/N $\sim5$ (see Figure 1 of \citealt{Mor2014}). On the bright side, VMC is limited by saturation at $K_{\rm s}\sim11.4$ mag.   The majority of RR Lyrae stars in our sample are located within the VMC tile LMC 5\_5. Observations of the tile LMC 5\_5 were performed in 15 epochs taken in the period from 2010, October 30, to 2012, January 11. For two epochs of observation the ellipticity was too high, so these data were not considered in the analysis. Among the remaining 13 epochs there are 11 deep and 2 shallow epochs. Since shallow observations were obtained in good seeing conditions their S/N was enough to detect the RR Lyrae stars. In the following analysis we used all 13 available epochs to fit the light curves of the RR Lyrae stars.  PSF photometry of the time-series data for this tile was performed on the homogenised epoch-tile images \citep{Rub2012} using the IRAF Daophot  packages \citep{Stet1990}. On each epoch-tile image the PSF model was created using 2500 stars uniformly distributed, finally the Daophot ALLSTAR routine was used to perform the PSF photometry on all epoch images and time-series catalogues were correlated within a tolerance of one arcsec.
 
\begin{figure}
\includegraphics[width=\linewidth]{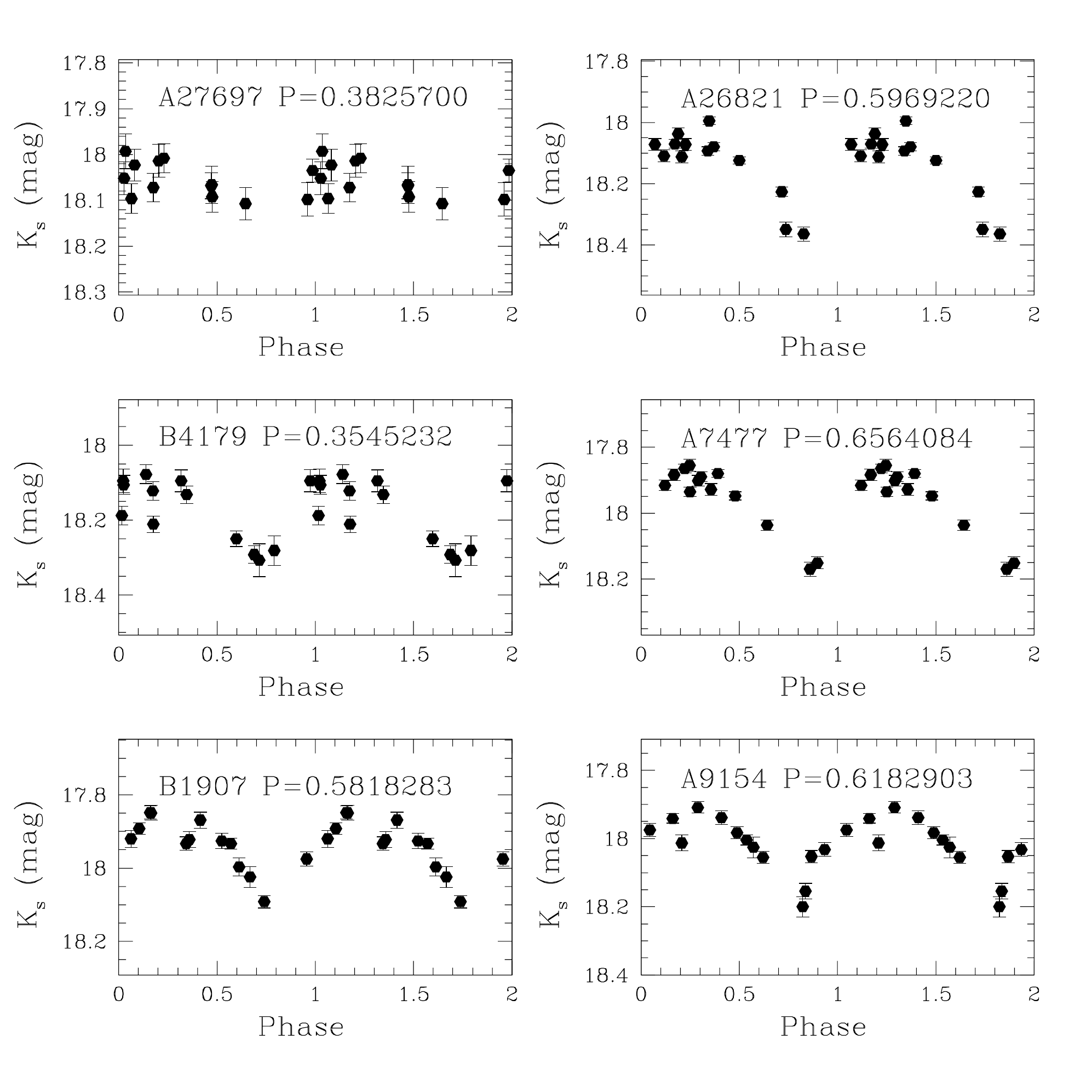}
\caption{Examples of  $K_{\rm s}$-band light curves for RR Lyrae stars in our sample. Identification numbers are from \citet{Fab2005}, periods are from the OGLE~III catalogue \citep{Sosz2009} and are given in days.}
\label{lc}
\end{figure}

We have cross-matched our sample of 82 RR Lyrae stars against the PSF photometry catalogue of the VMC tile LMC 5\_5. VMC counterparts for 71 objects were found within a pairing radius of $1\arcsec$.
Among them, 70 sources have 13 epochs in the $K_{\rm s}$-band, while for one object (B4749) we have observations only in 6 epochs. 
Six data points are not enough for a reliable fit of the  light curve and, consequently, for the robust determination of the mean $K_{\rm s}$ magnitude, hence, we discarded this source from the following analysis and proceeded with the 70 RR Lyrae stars, for which 13 epochs in the $K_{\rm s}$-band exist. We derived the mean $K_{\rm s}$  magnitudes of these 70 RR Lyrae stars by Fourier fitting the light curves with the GRaphical Analyzer of TImes Series package (GRATIS, custom software  developed at the Observatory of Bologna by P. Montegriffo, see e.g. \citealt{Clem2000}). To fit the light curves we discarded obvious outliers. Nevertheless, after the $\sigma$-clipping procedure, each source still has 11 or more data points. Examples of the $K_{\rm s}$ light curves are shown in Figure~\ref{lc}. 

After deriving $K_{\rm s}$ mean magnitudes 
we performed the dereddening procedure. \citet{Clem2003} estimated reddening values of $E(B-V)$ = $0.116\pm 0.017$ and $0.086\pm0.017$ mag in LMC field A and B, respectively, using the method from \citet{Stur1966} and the colours of the edges of the instability strip defined by the RR Lyrae variables. Applying the coefficients from \citet{Car1989} of $A_K/A_V$ = 0.114 and assuming a ratio of total to selective absorption of  $R_V$ = 3.1, we estimated the extinction in the $K_{\rm s}$-band as:
\begin{equation}
A_{K_{\rm s}}=0.35 \times E(B-V)
\end{equation}

Table~\ref{tab1} summarizes  the properties of the sample of 70 RR Lyrae stars which have a counterpart in the VMC catalogue. First and second columns give the  identification of the stars in \citet{Fab2005} and in the OGLE~III catalogue, respectively. The table also shows coordinates and the classification of the stars from the OGLE~III catalogue, metallicities with errors from \citet{Grat2004} and dereddened mean $K_{\rm s}$ magnitudes, determined with the GRATIS package, along with their errors.


\begin{deluxetable}{llccccccccc}
\tabletypesize{\scriptsize}
\tablecaption{Properties of the 70 RR Lyrae stars in the bar of the LMC analyzed in this paper\label{tab1}}
\tablewidth{460pt}
\tablehead{
\colhead{Star} & \colhead{OGLE ID} & \colhead{RA} & \colhead{DEC} & \colhead{Type} &
\colhead{[Fe/H]} & \colhead{$\sigma_{\rm [Fe/H]}$} & \colhead{P} &
\colhead{$\langle K_{\rm s,0}\rangle$} & \colhead{$\sigma_{\langle K_{\rm s,0}\rangle}$}\\
{}&{}&{(J2000)}&{(J2000)}&{}&{(dex)}&{(dex)}&{(days)}&{(mag)}&{(mag)}\\ 
}
\startdata
A28665 &OGLE-LMC-RRLYR-12944 &   5:22:06.55 &$-$70:27:55.6 &RRc	&   $-$0.63 &  0.24  &  0.3008299   &    18.450 & 0.046  \\
A7864  &OGLE-LMC-RRLYR-13857 &   5:23:39.25 &$-$70:31:38.1 &RRc	&   $-$1.36 &  0.22  &  0.3129458   &    18.550 & 0.055  \\
B4946  &OGLE-LMC-RRLYR-10621 &   5:18:11.08 &$-$70:59:35.6 &RRc	&   $-$1.11 &  0.25  &  0.3130142   &    18.394 & 0.054  \\
A2636  &OGLE-LMC-RRLYR-13548 &   5:23:09.09 &$-$70:39:08.1 &RRc	&   $-$1.61 &  0.29  &  0.3154437   &    18.562 & 0.045  \\
A8837  &OGLE-LMC-RRLYR-13326 &   5:22:45.70 &$-$70:30:14.3 &RRc	&   $-$1.52 &  0.22  &  0.3165579   &    18.660 & 0.089  \\
A8622  &OGLE-LMC-RRLYR-13164 &   5:22:28.93 &$-$70:30:35.9 &RRc	&   $-$1.44 &  0.28  &  0.3212334   &    18.426 & 0.032  \\
A7231  &OGLE-LMC-RRLYR-13680 &   5:23:22.42 &$-$70:32:35.4 &RRc	&   $-$1.46 &  0.26  &  0.3228047   &    18.236 & 0.051  \\
A2234  &OGLE-LMC-RRLYR-13479 &   5:23:01.47 &$-$70:39:44.4 &RRc	&   $-$1.53 &  0.18  &  0.3228060   &    18.292 & 0.044  \\
A4388  &OGLE-LMC-RRLYR-12614 &   5:21:31.67 &$-$70:36:46.3 &RRc	&   $-$1.33 &  0.27  &  0.3417737   &    18.411 & 0.048  \\
A10113 &OGLE-LMC-RRLYR-14046 &   5:24:00.38 &$-$70:28:06.1 &RRc	&   $-$1.52 &  0.25  &  0.3506618   &    18.197 & 0.036  \\
B6255  &OGLE-LMC-RRLYR-10111 &   5:17:17.88 &$-$70:57:26.4 &RRc	&   $-$1.52 &  0.16  &  0.3535596   &    18.305 & 0.038  \\
B4179  &OGLE-LMC-RRLYR-10142 &   5:17:19.95 &$-$71:01:02.1 &RRc	&   $-$1.53 &  0.27  &  0.3545232   &    18.150 & 0.034  \\
A8812  &OGLE-LMC-RRLYR-13150 &   5:22:26.44 &$-$70:30:19.1 &RRc	&   $-$1.23 &  0.24  &  0.3549660   &    18.281 & 0.036  \\
A26715 &OGLE-LMC-RRLYR-12593 &   5:21:29.33 &$-$70:29:23.4 &RRc	&   $-$1.39 &  0.18  &  0.3569006   &    18.308 & 0.032  \\
A2024  &OGLE-LMC-RRLYR-13572 &   5:23:11.02 &$-$70:40:03.3 &RRc	&   $-$1.62 &  0.26  &  0.3590534   &    18.246 & 0.043  \\
B6164  &OGLE-LMC-RRLYR-10612 &   5:18:10.17 &$-$70:57:30.7 &RRc	&   $-$1.88 &  0.22  &  0.3744821   &    18.039 & 0.045  \\
A27697 &OGLE-LMC-RRLYR-13012 &   5:22:14.03 &$-$70:28:35.0 &RRc	&   $-$1.33 &  0.25  &  0.3825700   &    18.030 & 0.023  \\
A19450 &OGLE-LMC-RRLYR-13841 &   5:23:37.95 &$-$70:34:06.7 &RRab&   $-$0.76 &  0.13  &  0.3979182   &    18.481 & 0.071  \\
B7064  &OGLE-LMC-RRLYR-10708 &   5:18:18.63 &$-$70:55:58.7 &RRc	&   $-$2.03 &  0.20  &  0.4004744   &    18.029 & 0.043  \\
B6957  &OGLE-LMC-RRLYR-10702 &   5:18:18.08 &$-$70:56:08.7 &RRc	&   $-$1.48 &  0.18  &  0.4047399   &    18.027 & 0.043  \\
B23502 &OGLE-LMC-RRLYR-10509 &   5:18:00.25 &$-$70:54:31.0 &RRab  &   $-$1.55 &  0.14  &  0.4724681   &    18.243 & 0.076  \\
A3061  &OGLE-LMC-RRLYR-13704 &   5:23:25.18 &$-$70:38:28.9 &RRab  &   $-$1.26 &  0.12  &  0.4744410   &    18.415 & 0.041  \\
B10811 &OGLE-LMC-RRLYR-10684 &   5:18:16.01 &$-$71:04:27.0 &RRab  &   $-$1.42 &  0.20  &  0.4760753   &    18.197 & 0.036  \\
B3400  &OGLE-LMC-RRLYR-10072 &   5:17:14.51 &$-$71:02:26.6 &RRab  &   $-$1.45 &  0.24  &  0.4852148   &    18.346 & 0.092  \\
A7325  &OGLE-LMC-RRLYR-13855 &   5:23:39.13 &$-$70:32:24.8 &RRab  &   $-$1.18 &  0.26  &  0.4864544   &    18.223 & 0.046  \\
B3033  &OGLE-LMC-RRLYR-10659 &   5:18:14.04 &$-$71:03:00.5 &RRab  &   $-$1.26 &  0.21  &  0.4986975   &    18.130 & 0.066  \\
B2055  &OGLE-LMC-RRLYR-10108 &   5:17:17.44 &$-$71:04:50.2 &RRab  &   $-$1.70 &  0.23  &  0.5207746   &    18.254 & 0.074  \\
A26525 &OGLE-LMC-RRLYR-12811 &   5:21:52.50 &$-$70:29:28.7 &RRab  &   $-$1.41 &  0.22  &  0.5225029   &    18.168 & 0.053  \\
A7211  &OGLE-LMC-RRLYR-13092 &   5:22:21.17 &$-$70:32:43.9 &RRab  &   $-$1.33 &  0.19  &  0.5226857   &    18.193 & 0.041  \\
A2767  &OGLE-LMC-RRLYR-13634 &   5:23:17.75 &$-$70:38:55.9 &RRab  &   $-$1.37 &  0.08  &  0.5325871   &    18.054 & 0.036  \\
B24089 &OGLE-LMC-RRLYR-10345 &   5:17:43.51 &$-$70:54:02.7 &RRab  &   $-$1.48 &  0.16  &  0.5580613   &    18.094 & 0.069  \\
A8788  &OGLE-LMC-RRLYR-13678 &   5:23:22.41 &$-$70:30:14.6 &RRab  &   $-$1.61 &  0.21  &  0.5591710   &    18.197 & 0.036  \\
A6398  &OGLE-LMC-RRLYR-13294 &   5:22:40.76 &$-$70:33:50.2 &RRab  &   $-$1.40 &  0.30  &  0.5619466   &    17.957 & 0.027  \\
A7247  &OGLE-LMC-RRLYR-13708 &   5:23:25.58 &$-$70:32:33.4 &RRab  &   $-$1.38 &  0.21  &  0.5621512   &    18.045 & 0.049  \\
A25301 &OGLE-LMC-RRLYR-12638 &   5:21:34.00 &$-$70:30:24.5 &RRab  &   $-$1.58 &  0.27  &  0.5631146   &    18.268 & 0.051  \\
A15387 &OGLE-LMC-RRLYR-12603 &   5:21:30.43 &$-$70:37:11.3 &RRab  &   $-$1.81 &  0.12  &  0.5635914   &    18.049 & 0.052  \\
B22917 &OGLE-LMC-RRLYR-10713 &   5:18:19.10 &$-$70:54:56.1 &RRab  &   $-$1.29 &  0.16  &  0.5646803   &    18.179 & 0.054  \\
A9245  &OGLE-LMC-RRLYR-13536 &   5:23:07.67 &$-$70:29:36.5 &RRab  &   $-$1.27 &  0.18  &  0.5678763   &    18.013 & 0.035  \\
A12896 &OGLE-LMC-RRLYR-13330 &   5:22:46.15 &$-$70:38:54.9 &RRab  &   $-$1.53 &  0.10  &  0.5719281   &    18.143 & 0.035  \\
A7609  &OGLE-LMC-RRLYR-13941 &   5:23:48.39 &$-$70:32:00.3 &RRab  &   $-$1.63 &  0.11  &  0.5724984   &    18.023 & 0.046  \\
B7442  &OGLE-LMC-RRLYR-10082 &   5:17:15.73 &$-$70:55:26.8 &RRab  &   $-$1.58 &  0.11  &  0.5740274   &    18.096 & 0.048  \\
A25362 &OGLE-LMC-RRLYR-13848 &   5:23:38.53 &$-$70:30:08.5 &RRab  &   $-$1.39 &  0.15  &  0.5787944   &    18.033 & 0.052  \\
B1907  &OGLE-LMC-RRLYR-10638 &   5:18:12.36 &$-$71:04:59.5 &RRab  &   $-$1.70 &  0.26  &  0.5818283   &    17.915 & 0.036  \\
A4974  &OGLE-LMC-RRLYR-13372 &   5:22:51.26 &$-$70:35:47.7 &RRab  &   $-$1.36 &  0.10  &  0.5820430   &    17.992 & 0.054  \\
B6798  &OGLE-LMC-RRLYR-10044 &   5:17:11.37 &$-$70:56:32.6 &RRab  &   $-$1.40 &  0.23  &  0.5822610   &    17.910 & 0.114  \\
B14449 &OGLE-LMC-RRLYR-09999 &   5:17:05.37 &$-$71:01:40.9 &RRab  &   $-$1.70 &  0.13  &  0.5822854   &    18.118 & 0.071  \\
A9494  &OGLE-LMC-RRLYR-13354 &   5:22:49.26 &$-$70:29:13.5 &RRab  &   $-$1.69 &  0.28  &  0.5844615   &    17.874 & 0.035  \\
A18314 &OGLE-LMC-RRLYR-13353 &   5:22:49.13 &$-$70:34:59.2 &RRab  &   $-$1.42 &  0.18  &  0.5875708   &    18.093 & 0.030  \\
A10487 &OGLE-LMC-RRLYR-13126 &   5:22:24.61 &$-$70:27:40.6 &RRab  &   $-$1.49 &  0.11  &  0.5909585   &    18.030 & 0.016  \\
A10214 &OGLE-LMC-RRLYR-12609 &   5:21:31.14 &$-$70:28:12.0 &RRab  &   $-$1.48 &  0.12  &  0.5918196   &    17.904 & 0.065  \\
A28066 &OGLE-LMC-RRLYR-13765 &   5:23:30.10 &$-$70:28:11.0 &RRab  &   $-$1.44 &  0.17  &  0.5959296   &    18.007 & 0.060  \\
A26821 &OGLE-LMC-RRLYR-12831 &   5:21:53.95 &$-$70:29:17.5 &RRab  &   $-$1.37 &  0.13  &  0.5969220   &    18.130 & 0.047  \\
B2249  &OGLE-LMC-RRLYR-10061 &   5:17:13.06 &$-$71:04:27.1 &RRab  &   $-$1.56 &  0.15  &  0.6030630   &    17.999 & 0.050  \\
A16249 &OGLE-LMC-RRLYR-12960 &   5:22:08.27 &$-$70:36:31.0 &RRab  &   $-$1.87 &  0.12  &  0.6067385   &    18.060 & 0.045  \\
A4933  &OGLE-LMC-RRLYR-13175 &   5:22:30.05 &$-$70:35:53.7 &RRab  &   $-$1.48 &  0.12  &  0.6134920   &    17.768 & 0.027  \\
A7734  &OGLE-LMC-RRLYR-12956 &   5:22:07.86 &$-$70:31:59.8 &RRab  &   $-$1.40 &  0.15  &  0.6149615   &    17.888 & 0.036  \\
A2525  &OGLE-LMC-RRLYR-13788 &   5:23:32.45 &$-$70:39:15.3 &RRab  &   $-$2.06 &  0.14  &  0.6161452   &    17.964 & 0.051  \\
A9154  &OGLE-LMC-RRLYR-13494 &   5:23:02.93 &$-$70:29:44.6 &RRab  &   $-$1.66 &  0.14  &  0.6182903   &    17.972 & 0.029  \\
B1408  &OGLE-LMC-RRLYR-10067 &   5:17:13.84 &$-$71:06:06.9 &RRab  &   $-$1.70 &  0.11  &  0.6297088   &    18.021 & 0.013  \\
A5589  &OGLE-LMC-RRLYR-12968 &   5:22:09.60 &$-$70:35:02.5 &RRab  &   $-$1.60 &  0.13  &  0.6375745   &    17.948 & 0.035  \\
A7468  &OGLE-LMC-RRLYR-13176 &   5:22:30.06 &$-$70:32:20.6 &RRab  &   $-$1.55 &  0.11  &  0.6386908   &    18.043 & 0.041  \\
A25510 &OGLE-LMC-RRLYR-13002 &   5:22:13.43 &$-$70:30:11.4 &RRab  &   $-$1.72 &  0.11  &  0.6495506   &    17.713 & 0.038  \\
A8720  &OGLE-LMC-RRLYR-13956 &   5:23:50.19 &$-$70:30:16.7 &RRab  &   $-$1.88 &  0.34  &  0.6508174   &    17.847 & 0.037  \\
B7063  &OGLE-LMC-RRLYR-10973 &   5:18:44.05 &$-$70:55:55.8 &RRab  &   $-$1.49 &  0.14  &  0.6548698   &    17.867 & 0.023  \\
B7620  &OGLE-LMC-RRLYR-10541 &   5:18:03.58 &$-$70:55:03.1 &RRab  &   $-$2.05 &  0.12  &  0.6561602   &    17.689 & 0.034  \\
A7477  &OGLE-LMC-RRLYR-14068 &   5:24:02.97 &$-$70:32:08.6 &RRab  &   $-$1.67 &  0.28  &  0.6564084   &    17.955 & 0.017  \\
A28293 &OGLE-LMC-RRLYR-12758 &   5:21:46.13 &$-$70:28:13.3 &RRab  &   $-$1.74 &  0.10  &  0.6602890   &    17.979 & 0.050  \\
A6426  &OGLE-LMC-RRLYR-13196 &   5:22:32.51 &$-$70:33:48.7 &RRab  &   $-$1.59 &  0.09  &  0.6622400   &    17.868 & 0.038  \\
A3948  &OGLE-LMC-RRLYR-13285 &   5:22:40.40 &$-$70:37:17.0 &RRab  &   $-$1.46 &  0.12  &  0.6623845   &    17.944 & 0.036  \\
A8094  &OGLE-LMC-RRLYR-13306 &   5:22:43.06 &$-$70:31:23.7 &RRab  &   $-$1.83 &  0.12  &  0.7420663   &    17.890 & 0.033  \\

\enddata

\tablecomments{Columns report: 1) Star identification from \citet{Fab2005}; 2) Identification from the OGLE~III catalogue \citep{Sosz2009}; 3) Right ascension (OGLE); 4) Declination (OGLE); 5) Type; 6) Metallicity from \citet{Grat2004}; 7) Metallicity error \citep{Grat2004}; 8) Period (OGLE); 9) Dereddened mean $K_{\rm s}$ magnitude from the VMC data, determined from the analysis of the light curve with GRATIS; 10) Error of  the mean $K_{\rm s}$ magnitude.} 

\end{deluxetable}

\section{${\rm PL_{K_{\rm s}}Z}$ relation of RR Lyrae stars\label{LMC_PL}}
\subsection{Method\label{meth}}
Using the dereddened mean $K_{\rm s}$ magnitudes of the 70 RR Lyrae stars  derived as described in Section~\ref{dat}, spectroscopically determined metallicities from \citet{Grat2004} and accurately estimated periods from the OGLE~III catalogue (with RRc stars "fundamentalized"  by adding 0.127 to the logarithm of the period) 
we can now fit the ${\rm PL_{K_{\rm s}}Z}$ relation. The fit is performed using a fitting approach developed specifically for this work.

Fitting a line to data is a common exercise in science. Most common approaches use Minimum-Least-Squares methods, however these are often based on assumptions which do not always hold for real observational data. The most basic methods assume that data are drawn from a thin line with errors, which are Gaussian, perfectly known, and exist in one axis only. These assumptions do not hold in the present case, as we have an unknown but potentially significant intrinsic dispersion, non-negligible errors in two dimensions ($K_{\rm s}$ and [Fe/H]), and the possibility of inaccuracy in the formal error estimates (e.g. in the determination of the precision metallicity estimates). 

We therefore follow the prescription of \citet{Hogg2010}, who develop a method for fitting a line to data which avoids the problems highlighted above by statistical modelling of the data. They present a method for use in two dimensions, which has been extended to three dimensions in this paper. 

The method assumes that the data is drawn from a plane of the form 
\begin{equation}
L{\rm s}(P,{\rm [Fe/H]}) = A\,{\rm logP} + B\,{\rm [Fe/H]} + C \quad
\end{equation}
where $A$ is the slope in the $\rm logP$ axis, $B$ is the slope in the metallicity ${\rm [Fe/H]}$ axis, and $C$ is the intercept. We assume a uniform Gaussian intrinsic dispersion around the luminosity axis, plus the scatter caused by the Gaussian observational errors. The exact mathematical definition is given in Appendix \ref{derivation}. The method utilises adaptive Markov Chain Monte Carlo (MCMC) methods \citep{emcee} to evaluate the posterior probability density function (PDF) of each parameter, given an input dataset, and returns the maximum a posteriori probability (MAP) estimate of each parameter, the formal error estimate, and the full posterior PDF. The formal error estimate is obtained from the 16\% and 84\% quartiles of the posterior PDF of the parameters, which give the 1$\sigma$ formal error estimate assuming that the posterior PDF is approximately Normal. The free fit parameters are: the slope in logP, the slope in metallicity, the zero-point, and the intrinsic dispersion perpendicular to the magnitude axis.

By applying this method we found the following relation between period, metallicity and mean apparent $K_{\rm s}$ magnitude:

\begin{eqnarray}\displaystyle \label{template}
K_{\rm s,0} 
& = & (-2.73 \pm 0.25) \textrm{log}P + (0.03 \pm 0.07){\rm [Fe/H]}\nonumber\\
& + & (17.43 \pm 0.01)
\end{eqnarray}

\begin{figure}
\includegraphics[width=10cm]{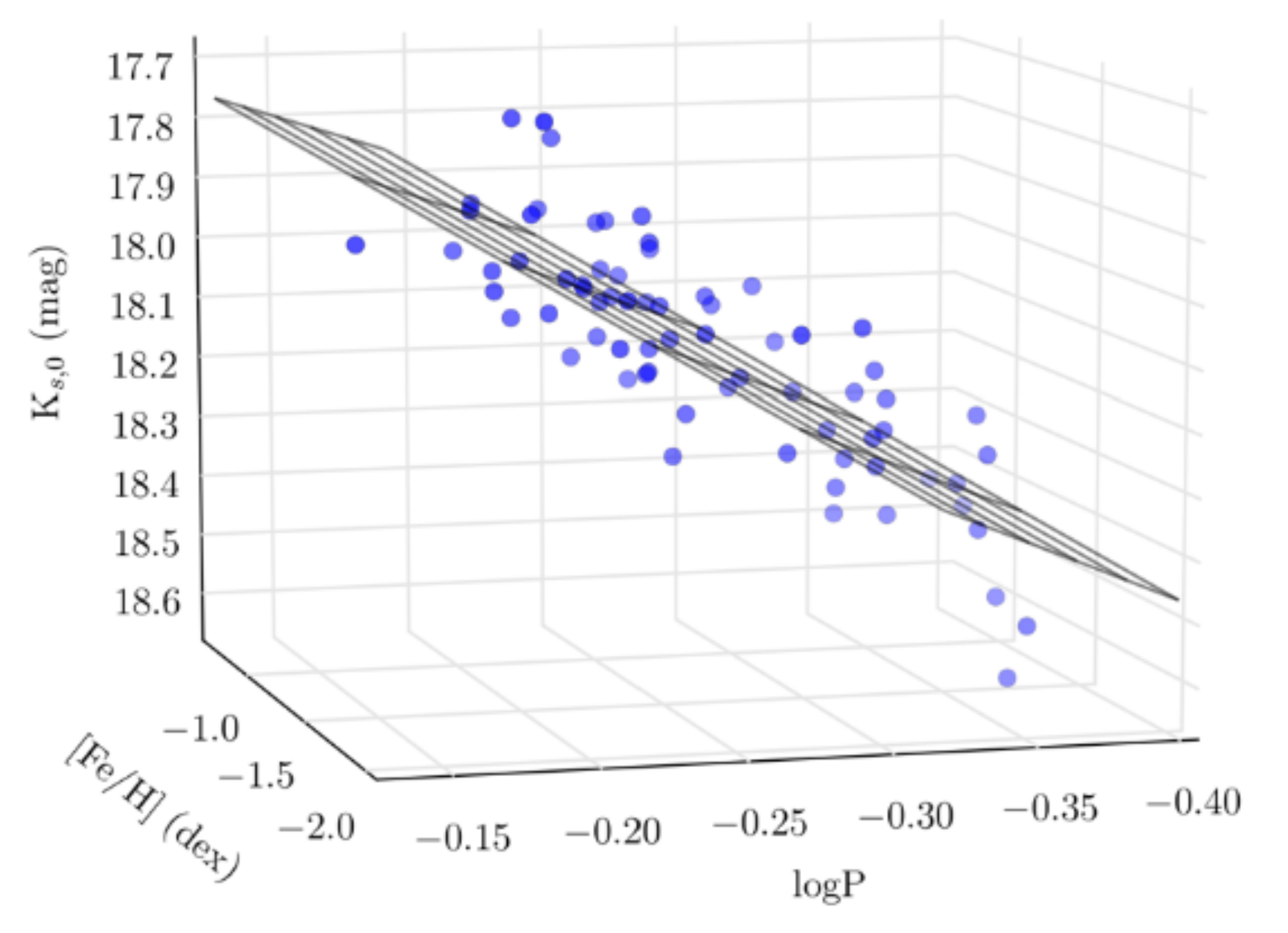}
\includegraphics[width=6.5cm]{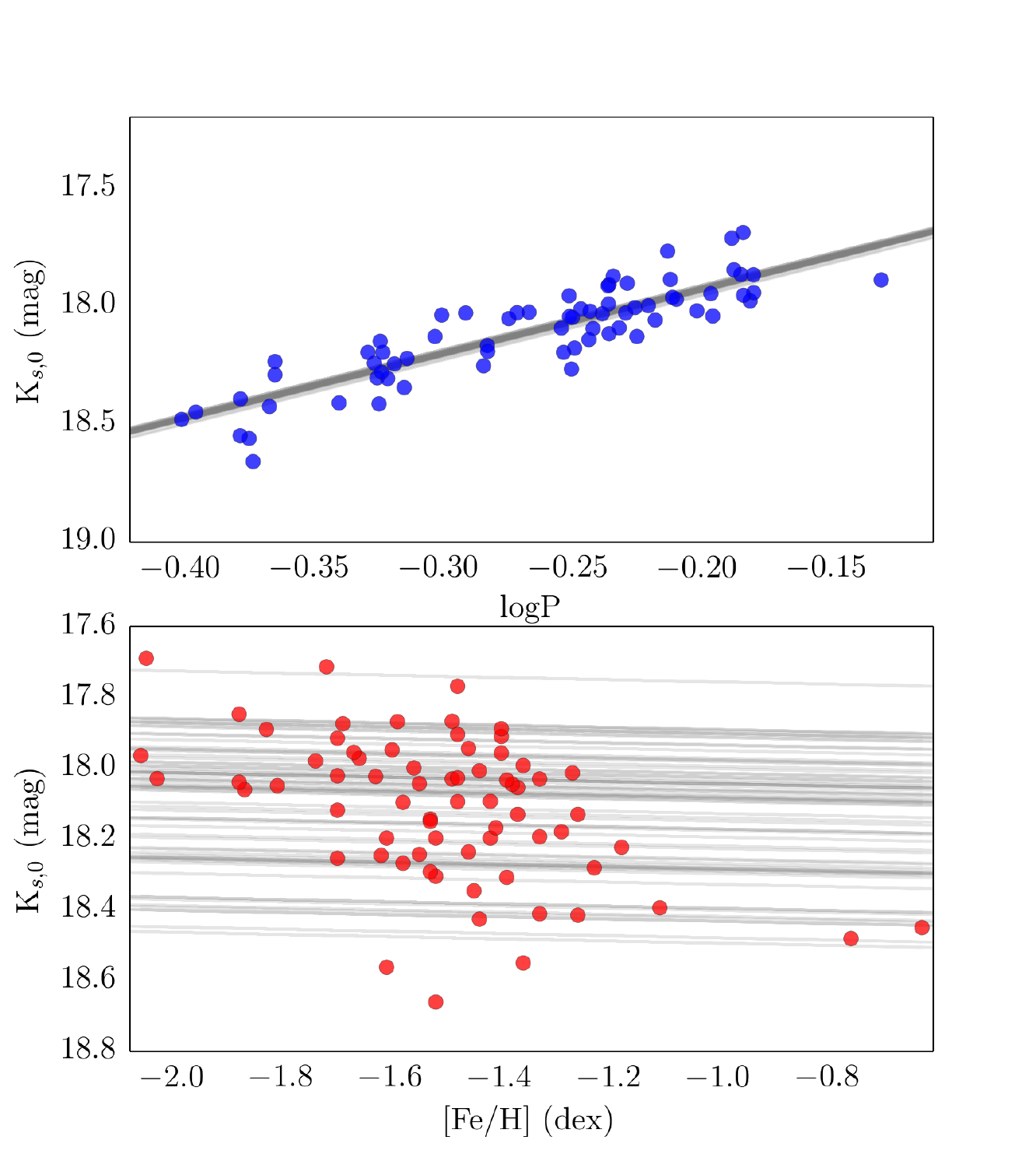}
\caption{{\it Left panel}: ${\rm PL_{K_{\rm s}}Z}$ relation of the 70 LMC RR Lyrae stars (Equation~\ref{template}) analyzed in the paper, in the period-luminosity-metallicity space. {\it Right panels}: Projections of the ${\rm PL_{K_{\rm s}}Z}$ relation (Equation~\ref{template}) on the ${{\rm log(P)}-K_{\rm s}}$ (top) and $K_{\rm s}-{\rm [Fe/H]}$ (bottom) planes. Grey lines represent lines of equal metallicity (top) and equal period (bottom). See text for the details.  Uncertainties in  the $K_{\rm s}$ magnitude and ${\rm [Fe/H]}$ are omitted to simplify the figure, but  they are provided in Table~\ref{tab1}.}
\label{PL_templ}
\end{figure}

The intrinsic dispersion of the relation is found to be  0.01 mag. The RMS deviation of the data around the relation, neglecting the intrinsic dispersion, is 0.1 mag. 
 Since the reddening in the $K_{\rm s}$ band is negligible we suggest that the effects of the LMC depth cause the intrinsic dispersion of the relation. 
 The left panel of  Figure~\ref{PL_templ} presents the ${\rm PL_{K_{\rm s}}Z}$ relation (Equation~\ref{template}) of the 70 LMC RR Lyrae stars in the period-luminosity-metallicity space, 
whereas the right panels show the projection of the ${\rm PL_{K_{\rm s}}Z}$ on the ${\rm log(P)}-K_{\rm s}$ (top-right panel) and $K_{\rm s}-{\rm [Fe/H]}$ (bottom-right panel) planes.
The grey lines in the figure are lines of equal metallicity (top-right) or equal period (bottom-right). The method finds the relation (values of A, B, and C for the relation $K_{\rm s}$ = A\,logP+B\,[Fe/H]+C) in the three dimensions (logP, $K_{\rm s}$, and [Fe/H]). Each of the grey lines in the top-right plot are therefore $K_{\rm s} =$A\,logP+B\,[Fe/H]+C for the full range of periods, at the metallicity of each star (one line per star). Thus, by following the line up and down it is seen how $K_{\rm s}$ changes with period at some specific metallicity. The lines do not always cross the points on the diagram because the line is the result of the fit, and the points are affected by errors and intrinsic dispersion so may be above or below the fit.  In the bottom-right plot the lines are $K_{\rm s}$ = A\,logP+B\,[Fe/H]+C for the full range of metallicity with logP taken from each star.

It is worth noting that we find a very small dependence of the $K_{\rm s}$ magnitude on metallicity. However, the metal abundance range spanned by the adopted sample does not reach the highest values (up to solar and supersolar) observed in the MW bulge and disk RR Lyrae populations. In order to study the effect of the adopted range of metallicities on the slope of the ${\rm PL_{K_{\rm s}}Z}$ relation  we derived this relation also for MW RR Lyrae stars. We discuss the results in Section~\ref{sec:MW}. 

\subsection{Zero-point of the ${\rm PL_{K_{\rm s}}Z}$ relation in the LMC\label{zero-p}}

To use the derived ${\rm PL_{K_{\rm s}}Z}$ relation for determining distances it is necessary to calibrate its zero-point. This can be done in a number of different ways. In this paper we follow two different approaches: the first one is based on adopting a value for the distance of the LMC; in the second one we use the absolute magnitudes of Galactic RR Lyrae stars for which trigonometric parallaxes have been measured with the {\it HST}/FGS. Both approaches have their advantages and disadvantages, we discuss them in the following  sections.

\subsubsection{Zero-point based on the LMC distance}
The LMC is widely considered the first rung of the cosmic distance ladder as it contains a large number of different distance indicators,  such as Cepheids, RR Lyrae variables, eclipsing binaries (EBs), red giant branch (RGB) stars, etc., allowing the galaxy distance to be determined  with several independent techniques. 
Figure~8 of \citet{Ben2002} shows an impressive summary of LMC distance moduli  published during  the ten years from 1992 to 2001. Values from 18.1 to 18.8 mag were reported in the  literature, with those smaller than 18.5 mag supporting the so-called "short" scale, and those larger than 18.5 mag, the "long" one.  In  more recent years the dramatic progress in the calibration of the different distance indicators has led the  dispersion in LMC distance moduli to shrink significantly. Extreme values such as those listed in \citet{Ben2002} are not very often  seen in the recent literature \citep{Clem2008}. Still a general consensus on the LMC distance  has not been fully reached yet.  
 Moreover, there have been significant concerns about a possible "publication bias" affecting the distance to the LMC (\citealt{Sch2008}, \citealt{Rub2012}, \citealt{Wal2012}). In particular, \citet{Sch2008}  claimed that from 2002 to 2007 June, 31 independent papers reported new measurements of the distance of the LMC, and the new values clustered tightly around the value $(m-M)_0 = 18.5 \pm 0.1$ mag, adopted by the  {\it HST} Key Project on the extragalactic distance scale (\citealt{Freed2001}). \citet{Sch2008} considered the effects of the "publication bias" to be the most likely cause of the clustering of LMC distance measurements. 
 
 A number of studies on the compilation of distances to the LMC as derived from different distance indicators can be found in the literature of the last 15 years (e.g., \citealt{Gib2000}; \citealt{Ben2002}; \citealt{Clem2003}; \citealt{Sch2008}; \citealt{deGr2014}). \citet{Clem2003} analysed the distance to the LMC measured using Population I and Population II standard candles and showed that all distance determinations converge within $1\sigma$ error on a distance modulus $(m-M)_0 = 18.515 \pm 0.085$ mag. 
The most recent compilation of  LMC distance moduli is that of 
\citet{deGr2014} who compiled 233 separate distance determinations, published from 1990 March until 2013 December,  and concluded that the canonical distance modulus of $(m-M)_0 = 18.49 \pm 0.09$ mag may be used for all practical purposes. 
 The compilation of \citet{deGr2014} includes the 
distance modulus 
of  $(m-M)_0 = 18.46 \pm 0.03$ estimated by \citet{Ripepi2012b} using LMC classical  Cepheids observed by the VMC survey, and the recent determination of  direct distances to eight long-period EBs in the LMC  
 by \citet{Pietr2013},  which is  claimed to be accurate to within $\sim2\%$: 
 $D_{LMC}$ = 49.97 $\pm$0.19 (stat) $\pm$ 1.11 (syst) kpc, corresponding to the distance modulus $(m-M)_0$= 18.493 $\pm$ 0.008 (stat) $\pm$  0.047 (syst) mag.  Furthermore, the model fitting of the light curves of different classes of pulsating stars in the LMC,  also based on different samples and hydrodynamical codes, provides values consistent with 18.5 mag (see \citealt{Bono2002};  \citealt{Marc2005}; \citealt{Kel2002,Kel2006}; \citealt{McN2007}).

The RR Lyrae stars in our  sample are located in a relatively small area close to the center of the LMC bar.  Neglecting depth/projection effects  they can be considered as being all at the same distance from us and close  to late-type EBs, which are all  located relatively close to the barycenter of the LMC as derived by \citet{Pietr2013}. Therefore, in the following we adopt for the distance modulus of the LMC the value published by \citet{Pietr2013}. We subtracted this value from the dereddened mean $K_{\rm s}$ apparent magnitudes of our 70 RR Lyrae stars and derived  absolute magnitudes in the $K_{\rm s}$  band  ($M_{K_{\rm s}}$). Then by applying the technique described in Section~\ref{meth} we derived  the relation between $K_{\rm s}$-band absolute magnitudes, periods and metallicities, with the zero-point entirely based on  the distance  to the LMC  by  \citet{Pietr2013}:
\begin{eqnarray}\displaystyle \label{M_LMC}
M_{K_{\rm s}} 
& = & (-2.73 \pm 0.25) \textrm{logP} + (0.03 \pm 0.07){\rm [Fe/H]}\nonumber\\
& - & (1.06 \pm 0.01)
\end{eqnarray}
 

\subsubsection{Zero-point based on trigonometric parallaxes of Galactic RR Lyrae stars\label{sec:ZP_Ben}}

In order to obtain an estimate of the ${\rm PL_{K_{\rm s}}Z}$ relation zero-point which is independent of the distance to the LMC and, in turn, be able to measure the  distance to this galaxy 
from the RR Lyrae ${\rm PL_{K_{\rm s}}Z}$ relation,
it is necessary to know the absolute magnitude of  the RR Lyrae stars with good accuracy. Trigonometric parallaxes remain the only direct method to measure distances and hence derive absolute magnitudes. \citet{Ben2011} derived absolute trigonometric parallaxes for five Galactic RR Lyrae stars (RZ Cep, XZ Cyg, SU Dra, RR Lyr and UV Oct) with the {\it HST}/FGS. With these parallaxes the authors estimated absolute magnitudes in the $K_{\rm s}$ and {\it V} passbands, corrected for interstellar extinction and Lutz-Kelker-Hanson bias (\citealt{LK1973}, \citealt{Han1979}). Absolute magnitudes in the $K_{\rm s}$-band, periods and metallicities from \citet{Ben2011}, and the slopes of the relation derived  in Equation~\ref{template} 
were used in order to determine a zero-point from each of these five MW RR Lyrae stars. The metallicities in \citet{Ben2011} are in the Zinn \& West metallicity scale and were converted to the metallicity scale in \citet{Grat2004} by adding 0.06 dex (see Section~\ref{dat}). The logarithm of the period of the RRc star RZ Cep was "fundamentalized" by adding 0.127. Then we calculated the weighted mean of the five zero-points, this corresponds to:  $-1.27\pm0.08$ mag. The relation between absolute magnitudes, periods and metallicities with the zero-point based on the five MW RR Lyrae stars from \citet{Ben2011} is:

\begin{eqnarray}\displaystyle \label{M_Ben}
M_{K_{\rm s}} 
& = & (-2.73 \pm 0.25) \textrm{logP} + (0.03 \pm 0.07){\rm [Fe/H]}\nonumber\\
& - & (1.27 \pm 0.08)
\end{eqnarray}

  A recent analysis (Monson 2015, private communication) shows that there is likely a typo in \citet{Ben2011} parallax for the RR Lyrae star RZ Cep. Hence, we excluded this star from the sample and derived the zero-point based on parallaxes of remaining four RR Lyrae stars (XZ Cyg, UV Oct, SU Dra and RR Lyr):

\begin{eqnarray}\displaystyle \label{M_Ben_RZout}
M_{K_{\rm s}} 
& = & (-2.73 \pm 0.25) \textrm{logP} + (0.03 \pm 0.07){\rm [Fe/H]}\nonumber\\
& - & (1.25 \pm 0.06)
\end{eqnarray}

Situation improves, however there is still a difference of $\sim0.2$ mag between the two zero-point obtained based on the distance to the LMC (Eq.~\ref{M_LMC}) and the one based on the {\it HST} parallaxes of four RR Lyrae stars (Eq.~\ref{M_Ben_RZout}). In fact, if we apply our ${\rm PL_{K_{\rm s}}Z}$ relation with zero-point calibrated on  \citet{Ben2011} parallaxes (Eq.~\ref{M_Ben_RZout})
to determine the absolute magnitudes of the 70 RR Lyrae stars in our sample,  we obtain a distance modulus for the LMC of  $(m-M)_0 =18.68\pm0.10$ mag. This distance modulus is about 0.2 mag longer than the  widely adopted value of  $(m-M)_0=18.5$ mag.

There are a number of possible explanations for the discrepancy between zero-points. First of all, we should remember that \citet{Pietr2013} results have been called into question by \citet{Sch2013} who,  in addition to concerns regarding possible  bandwagon effects, also pointed out that
\citet{Pietr2013} distance to the LMC 
differs significantly from the average distance inferred from four hot, early-type
EBs, D=47.1$\pm$1.4 kpc ($(m-M)_0 = 18.365 \pm 0.065$ mag), published by \citet{Guinan1998}, \citet{Fitzpatrick2002,Fitzpatrick2003}, and \citet{Ribas2002}.
Furthermore, in using the  late-type EBs to calibrate the RR Lyrae ${\rm PL_{K_{\rm s}}Z}$ relation we have implicitly assumed that  RR Lyrae stars and EBs are at same distance from us. 
 However,  when pushing for distance comparisons at a few percent level the effects of sample size, spatial distribution, depth and geometric  projection become important and properly accounting for the internal structure of the LMC may become necessary.  The RR Lyrae stars in our sample could be distributed along the whole depth of the LMC. Moreover, RR Lyrae stars and EBs from \citet{Pietr2013} could reside in different sub-structures of the LMC, which could be the reason for the systematic error in the determination of the zero-point (see e.g. \citealt{Mor2014}  for different features of the LMC structure traced by classical Cepheids, RR Lyrae stars and hot EBs). 
 
 On the other hand, when calibrating the zero-point by applying parallaxes of the  four MW RR Lyrae stars by  \citet{Ben2011} we implicitly assumed that the ${\rm PL_{K_{\rm s}}Z}$ relation is the same in the MW and in the LMC, which may not be true (see Subsection~\ref{sec:MW}). We may also wonder whether there might be unknown systematic errors affecting Benedict et al.'s parallaxes. These come from {\it HST} fields, which provide relative and not absolute trigonometric parallaxes. Absolute parallaxes of the reference stars in each field are estimated via a complex procedure of fitting the spectral type and luminosity class of  each star. A general formal error of 0.5 mas is applied to the absolute parallax of the reference stars, equal for all stars in all fields, and without justification. This could result in miscalculated estimates of the precision of the final absolute parallax measurements of the four RR Lyrae stars. The Lutz-Kelker bias is corrected a posteriori.
In this respect it is worth of notice that, according to \citet{Leeuw2007}, Hipparcos parallax of RR Lyrae itself, the only RR Lyrae variable for which the satellite measured the parallax with high precision ($\pm$ 0.64 mas), is about 0.31 mas smaller than \citet{Ben2011}'s parallax for the star, although consistent with it within the errors, hence, the corresponding distance modulus is about 0.17 mag longer. In any case, a great contribution to the determination of the zero-point of the RR Lyrae ${\rm PL_{K_{\rm s}}Z}$ relation is expected from the ESA astrometric satellite Gaia.  We discuss this topic in Section~\ref{sec:gaia}.

\subsection{${\rm PL_{K_{\rm s}}Z}$ relation of RR Lyrae stars in the MW\label{sec:MW}}

In spite of many studies in the literature, it remains still unsettled whether the RR Lyrae  ${\rm PL_{K_{\rm s}}Z}$   is a universal relation.  To investigate this matter we have derived the  ${\rm PL_{K_{\rm s}}Z}$  relation for RR Lyrae stars in the MW and compared it with the relation we have obtained in Section~\ref{zero-p} for the LMC variables. To this end we selected 23 MW RR Lyrae stars which have their absolute magnitudes known from Baade-Wesselink (hereinafter B-W) studies based on near-infrared data (\citealt{Jon1988, Jon1992, Fern1990, LJ1990, Cacc1992, Ski1993, Fern1994}, and references therein) and metallicities determined from abundance analysis of high resolution spectra. 
Information about these 23 RR Lyrae stars is presented in Table~\ref{tab:MW}. Star's coordinates in the table are from the SIMBAD database; periods, apparent $V$ and $K$ magnitudes and reddening E(B-V) are from \citet{Fern1998a}. The sample contains two first-overtone mode RR Lyrae stars (namely, T Sex and TV Boo). As done for  the LMC RRc stars,  their periods were fundamentalized by adding  +0.127 to the logarithm of the period. Absolute  $M_V$ and $M_K$ magnitudes  in  Columns 10 and 12  were taken from the  compilations of B-W results in Table~11 of \citet{Cacc1992}  and  from Table~16 of 
\citet{Ski1993} for the variable stars: WY Ant, W Crt and RV Oct.  According to \citet{Cacc1992} the $K$ magnitudes  of the stars analyzed with the B-W method are in the Johnson photometric system.
 Following the discussion in  \citet{Cacc1992} and \citet{Ski1993} we retained only 23  of the original lists of 29 stars field RR Lyrae stars  analysed with the B-W method, as  stars which are likely to be evolved (DX Del, SU Dra, SS Leo, BB Pup and W Tuc) were discarded. We also discarded DH Peg as there is suspect the star is a dwarf Cepheid (see \citealt{Feast2008} and discussion therein).  Furthermore, 
 following  \citet{Fern1994},  original $M_V$ and $M_K$  values were revised (i) assuming for  the  $p$  factor used  to convert the observed pulsation velocity to true pulsation velocity in B-W analyses the value  of $p$=1.38, and (ii) multiple determinations of individual stars were averaged.

 Metal abundances with related errors are needed to apply our fitting approach.
Several different spectroscopic studies have targeted the stars in  Table~\ref{tab:MW}.  In  Appendix B we provide a summary of  their  results. 
The largest and most homogeneous samples are those by  \citet{Clem1995} and \citet{Lambert1996}.  These Authors 
measured [Fe/H] abundances from 
high resolution spectra for several of the stars in Table~\ref{tab:MW} 
 and provided recalibrations of the $\Delta S$ index \citep{Prest1959}, from which metal abundances can be derived for the stars which lack abundance analysis.  For sake of homogeneity and ease of use in this paper we adopt metallicities and metallicities errors for the MW RR Lyrae stars as they are listed, ready for use, in Table~21 of 
 \citet{Clem1995}.  These  [Fe/H]  values  are the average of the FeI and FeII abundances,  adopting log $\epsilon$(FeI)= 7.56 and log $\epsilon$(FeII)=7.50 for the sun. They  are reported  along with the related errors 
  in columns 5 and 6 of  Table~\ref{tab:MW}. Metallicities in \citet{Lambert1996}  were obtained from the FeII abundance  and adopting log $\epsilon$(FeI)= 7.51 for the sun. 
Corresponding  [Fe/H]  values and errors, 
are reported in columns 2  and 3 of Table 4 in Appendix B. In  Appendix B we also present the  $PLZ$ relation obtained using  \citet{Lambert1996}'s metallicities and our approach.
By applying our fitting approach (Section~\ref{meth}) to the 23 MW RR Lyrae stars we derived the following  ${\rm PL_{K}Z}$ relation:

\begin{eqnarray}\displaystyle \label{M_MW}
M_{K} 
& = & (-2.53 \pm 0.36) \textrm{logP} + (0.07 \pm 0.04){\rm [Fe/H]}\nonumber\\
& - & (0.95 \pm 0.14)
\end{eqnarray}
 
The intrinsic dispersion of the relation is found to be  0.007 mag. The RMS deviation of the data around the relation, neglecting the intrinsic dispersion, is 0.086 mag. 
It should be noted that the metallicities listed in Table~\ref{tab:MW} may differ slightly from the metal abundances used in the B-W analysis of these stars. However, this is not of great concern as the B-W based on near-infrared data is mildly affected by small changes in metallicity and reddening.  
We also point out  that the rather large error of the logP term in Equation~\ref{M_MW} is largely driven by the large errors (0.15-0.25 mag)  in the $K$-band absolute magnitudes from the B-W analyses. 
This is confirmed by the exercise with Gaia simulated parallaxes we present in Section~4.
 
The slope in [Fe/H]  in Equation~\ref{M_MW} is higher than the slope obtained for the LMC RR Lyrae stars (Equations~\ref{M_LMC}, \ref{M_Ben_RZout}), although they are still consistent within the respective errors.  Equation~\ref{M_MW} was derived over a wide range of metallicities [-2.5; 0.17] dex, nevertheless  the slope of the metallicity term remains rather small.
 Thus, the relatively small metallicity range spanned by the LMC variables 
could be not responsible for the negligible dependence on metallicity of the RR Lyrae  ${\rm PL_{K_{\rm s}}Z}$ relation in the LMC.
 
\begin{figure}
\includegraphics[width=10cm]{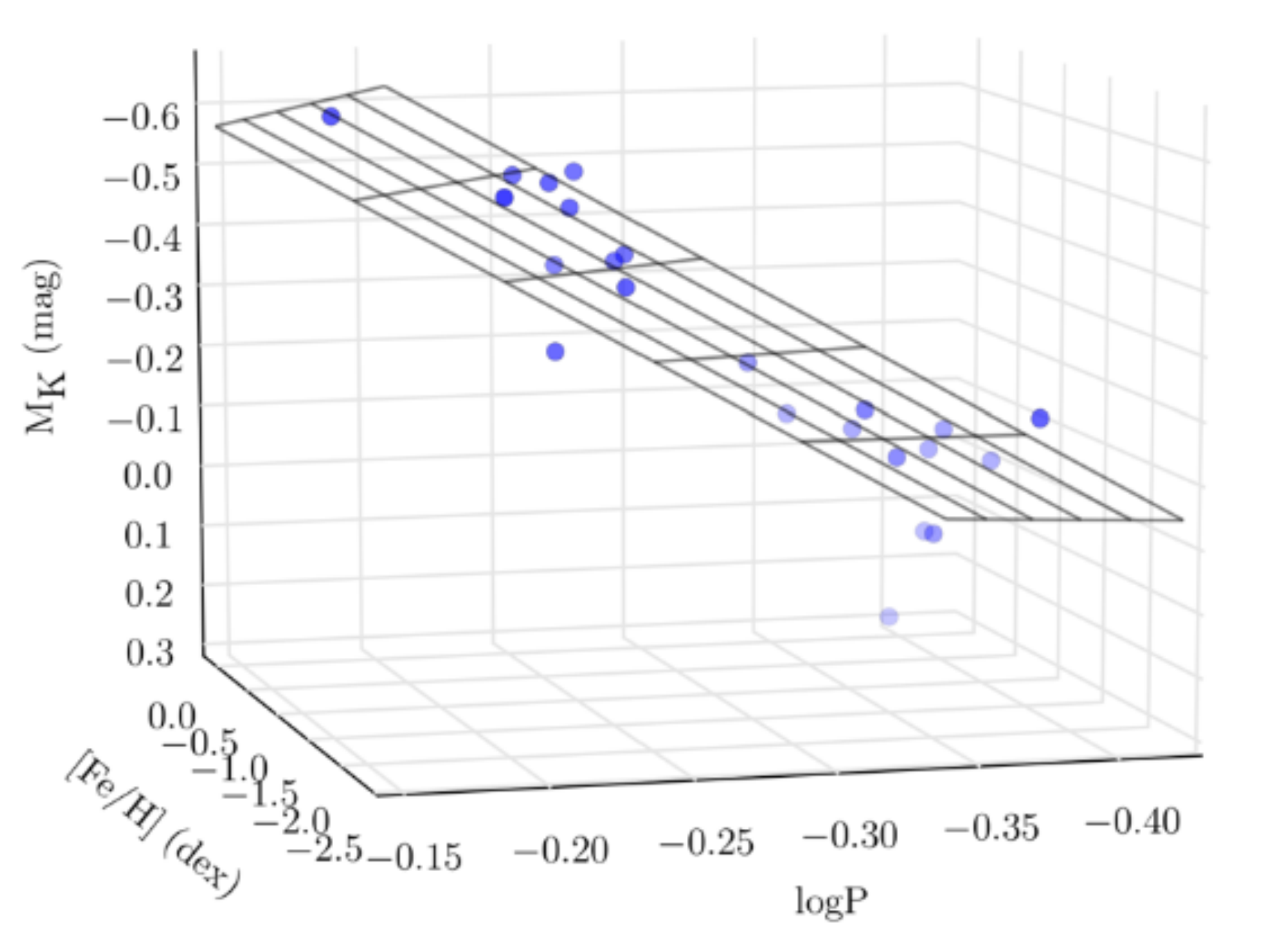}
\includegraphics[width=6.5cm]{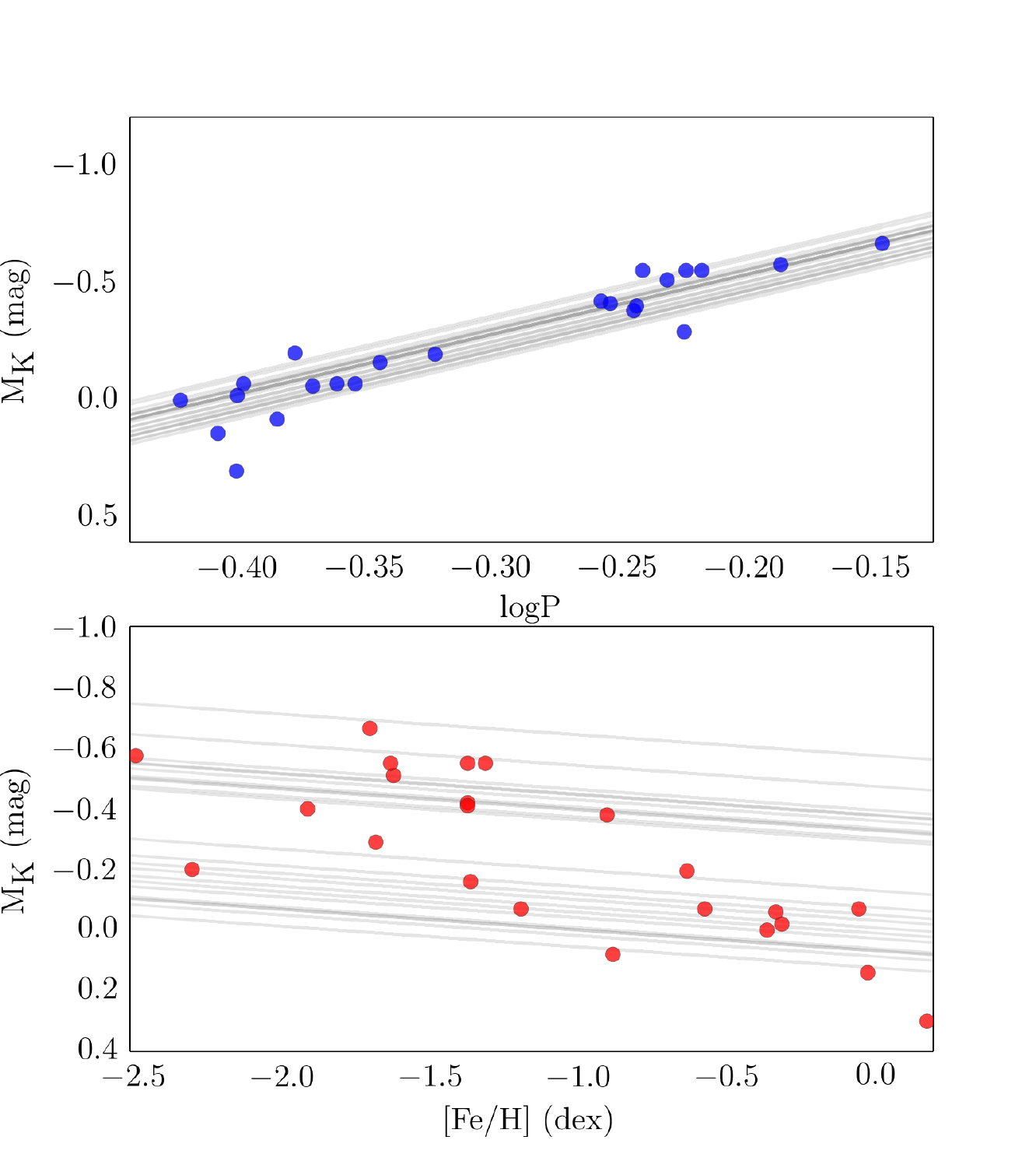}
\caption{{\it Left panel}: ${\rm PM_{K}Z}$ relation of the 23 MW RR Lyrae stars (Equation~\ref{M_MW}) in the period-luminosity-metallicity space. {\it Right panel}: Projections of the ${\rm PM_{K}Z}$ relation (Equation~\ref{M_MW}) on the ${\rm log(P)}-M_{K}$ (top panel) and $M_{K}-{\rm [Fe/H]}$ (bottom panel) planes. Grey lines represent lines of equal metallicities (top panel) and periods (bottom panel). See text for the details.  Uncertainties in the $M_{K}$ magnitude and ${\rm [Fe/H]}$ are omitted to simplify the figure, but  they are provided in Table~\ref{tab:MW}.}
\label{PL_MW}
\end{figure}

The distribution of the 23 MW RR Lyrae stars in the period-luminosity-metallicity space and the projections of the ${\rm PL_{K}Z}$ relation (Equation~\ref{M_MW}) on the ${\rm log(P)}-M_{K}$ and $M_{K}-{\rm [Fe/H]}$ planes are shown in Figure~\ref{PL_MW}. Grey lines are the same as in Figure~\ref{PL_templ} and are described in Section~\ref{meth}. 

Some concern may arise since the $K$ magnitudes of the 23 MW RR Lyrae stars used to derive Equation~\ref{M_MW} are in the Johnson photometric system (see, 
\citealt{Cacc1992}), whereas for the LMC RR Lyrae variables we have ${K_{\rm s}}$ photometry in the VISTA system\footnote{The VISTA system is tied to the 2MASS photometry (\citealt{Skrut2006}), with the difference in ${K_{\rm s}}$  magnitude only mildly depending on the $J - K_{\rm s}$ colour, and being  of the order of 3-4 mmag for the typical $J - K_{\rm s}$ colour of RR Lyrae stars.}. 
To address this issue we have reported in column  10 of Table~\ref{tab:MW}  the average ${K_{\rm s}}$ magnitudes in the 2MASS system of the 23 MW RR Lyrae stars as provided by \citet{Feast2008}. The difference 
with the Johnson $K$ average magnitudes listed in column  9 is very small (of the order of about 0.03 mag, on average) and definitely much smaller than individual errors in the B-W $K$-band absolute magnitudes of the MW variables  (0.15-0.25 mag),  or   errors in the ${K_{\rm s}}$ average apparent magnitudes of the LMC RR Lyrae stars (see column 10 of Table~1). Hence, we are confident that the difference in photometric system does  not affect significantly our comparison. 
%

\begin{deluxetable}{lrrccclrrrclc}
\tabletypesize{\footnotesize}
\tablecaption{Properties of 23 bright RR Lyrae stars in the Milky Way\label{tab:MW}}
\tablewidth{520pt}
\tablehead{
\colhead{Star} & \colhead{~~~~RA} &  \colhead{~~~~~DEC}  & \colhead{P} & \colhead{[Fe/H]} & \colhead{$\sigma_{\rm [Fe/H]}$}&  \colhead{E(B-V)}&
\colhead{$V$}& \colhead{$K$} &  \colhead{$K_{\rm s}$} & \colhead{$M_V$} &  \colhead{$M_K$} & \colhead{$\sigma_{M_V,M_K}$}\\
{}&{(deg)~~~~}&{(deg)~~~~}&{(day)~~~}&{(dex)~~}&{(dex)~~}&{~(mag)}&{~(mag)}&{(mag)}&{(mag)}&{(mag)}&{(mag)}&{(mag)}\\
}
\startdata

UU Cet  &   1.02135&$-$16.99764  &  0.606081 &  $-$1.38	&   0.08  &   0.015  &   12.08 &  10.85 &   10.837& 0.62  &     $-$0.55   &	0.15 \\
SW And  &   5.92954  & 29.40101   &  0.442279 &  $-$0.06	&   0.08  &   0.045  &   9.71  &  8.54  &  8.505 & 0.94  &      $-$0.07   &	0.15 \\
RR Cet  &  23.03405  &  1.34173    &  0.553025 &  $-$1.38	&   0.08  &   0.015  &   9.73  &  8.56  & 8.520 &  0.68  &      $-$0.42   &	0.15 \\
X Ari	&  47.12869  &  10.44590   &  0.651139 &  $-$2.50	         &   0.08  &   0.18   &   9.57   &  7.95  & 7.941 &   0.57  &      $-$0.575  &	0.15\\ 
AR Per  &  64.32165  & 47.40018   &  0.425549 &  $-$0.34       &   0.16  &   0.31   &   10.51 &  8.66  & 8.642 &  0.87  &      $-$0.06   &	0.25\\ 
RX Eri  &  72.43455&$-$15.74118  &  0.587246 &  $-$1.63	&   0.16  &   0.03   &   9.69  &  8.42  &   8.429 & 0.66  &     $-$0.51   &	0.25\\ 
RR Gem  & 110.38971 &  30.88318   &  0.397316 &  $-$0.32	&   0.16  &   0.075  &   11.38 &  10.26 & 10.275 &  0.89  &   $-$0.02   &	0.25 \\
TT Lyn  & 135.78245 &  44.58559   &  0.597438 &  $-$1.64	&   0.16  &   0.015  &   9.86  &  8.65  &  8.611 & 0.65  &     $-$0.55   &	0.15\\
T Sex	& 148.36833 &   2.05732    &  0.324698 &  $-$1.20	&   0.16  &   0.015  &   10.04 &  9.18  & 9.200 &  0.66  &    $-$0.07   &	0.25 \\
RR Leo  & 151.93108 &  23.99176   &  0.452387 &  $-$1.37	&   0.16  &   0.015  &   10.73 &  9.70	& 9.730 &   0.76  &   $-$0.16   &	0.15 \\
WY Ant  & 154.02061&$-$29.72845  &  0.574312 &  $-$1.32	&   0.16  &   0.06   &   10.87 &  9.64  &   9.674 & 0.55  &    $-$0.55   &	0.15 \\
W Crt	& 171.62351&$-$17.91435  &  0.412013 &  $-$0.89	&   0.16  &   0.03   &   11.54 &  10.56 & 10.539 &  0.96  &    +0.08    &	0.15 \\
TU UMa  & 172.45205 &  30.06733   &  0.557659 &  $-$1.38	&   0.16  &   0.015  &   9.82  &  8.67  &  8.660 &  0.70   &   $-$0.41   &	0.15 \\
UU Vir  & 182.14613& $-$0.45676   &  0.475606 &  $-$0.64	&   0.16  &   0.015  &   10.56 &  9.51  & 9.414 &   0.80   &    $-$0.195  & 0.15 \\
SW Dra  & 184.44429 &  69.51062   &  0.569670 &  $-$0.91	&   0.16  &   0.015  &   10.48 &  9.33  & 9.319 &   0.68  &       $-$0.38   &	0.15 \\
RV Oct  & 206.63230&$-$84.40177  &  0.571130 &  $-$1.92	&   0.16  &   0.09   &   10.98 &  9.51  & 9.526 &  0.68  &       $-$0.40    &	0.15 \\
TV Boo  & 214.15242 &  42.35992   &  0.312559 &  $-$2.31	&   0.16  &   0.015  &   10.97 &  10.22 & 10.248 &   0.58  &    $-$0.20    &	0.25 \\
RS Boo  & 218.38839 &  31.75462   &  0.377337 &  $-$0.37	&   0.16  &   0.015  &   10.37 &  9.45  & 9.507 &   0.85  &   +0.00     &	0.25 \\
VY Ser  & 232.75803 &   1.68382    &  0.714094 &  $-$1.71	&   0.08  &   0.03   &   10.13 &  8.78  & 8.826 &  0.61  &    $-$0.665  &	0.25 \\
V445 Oph& 246.17171 &$-$6.54165   &  0.397023 &  +0.17	&   0.08  &   0.195  &   11.05 &  9.24  &  9.262 & 1.09  &    +0.30     &	0.25 \\
TW Her  & 268.63000 &  30.41048   &  0.399601 &  $-$0.58	&   0.16  &   0.06   &   11.28 &  10.22 &  10.239 &  0.80   &   $-$0.07   &	0.15 \\
AV Peg  & 328.01164 &  22.57483   &  0.390380 &  $-$0.03	&   0.16  &   0.06   &   10.50  &  9.36  &  9.346 & 1.10   &     +0.14    &	0.15 \\
RV Phe  & 352.13106&$-$47.45362  &  0.596413 &  $-$1.69	&   0.16  &   0.015  &   11.94 &  10.72 &  10.768 & 0.86  &    $-$0.29   &	0.25 \\

\enddata

\tablecomments{The columns report: 1) Name of the star; 2) Right Ascension (J2000) from SIMBAD database; 3) Declination (J2000) from SIMBAD database; 4) Period from \citet{Fern1998a}; 5) Metallicity from \citet{Clem1995}; 6) Error in metallicity from \citet{Clem1995}; 7) Reddening from \citet{Fern1998a}; 8) $V$ magnitude from \citet{Fern1998a}; 9) $K$  magnitude in the Johnson system from \citet{Fern1998a};  10)  $K_{\rm s}$  magnitude in the 2MASS system from \citet{Feast2008}; 11) Absolute magnitude in the $V$ passband from \citet{Fern1994}; 
12) Absolute magnitude in the $K$ passband obtained from B-W analyses and corrected to p=1.38 (see text for details); 13) Errors in the absolute $V$, $K$ magnitudes. }

\end{deluxetable}

\subsection{Comparison with the literature}

The near-infrared ${\rm PL_{K}Z}$ relation of the RR Lyrae stars has been studied by several authors both from a theoretical and an observational point of view. \citet{Long1986} pioneering work was followed by \citet{LJ1990}, \citet{Ski1993} and  \citet{Jon1996}. A comprehensive analysis of the IR properties of RR Lyrae stars was performed by \citet{Nem1994}.

Some of the RR Lyrae ${\rm PL_{K}Z}$ relations available in the literature are presented in Table~\ref{ZP}. \citet{Bono2003} derived the semi-theoretical relation presented in the first row of Table~\ref{ZP}.  This theoretical relation has been derived from an extended set of RR 
Lyrae nonlinear hydrodynamical models spanning a wide range of chemical compositions (Z from 0.0001 to 0.02, which approximately corresponds to [Fe/H] from -2.45 to -0.15 dex). \citet{Cat2004} presented a theoretical calibration of the RR Lyrae ${\rm PL_{K}Z}$ relation based on synthetic HB models computed for several different metallicities, fully taking into
account evolutionary effects besides the effect of chemical composition. They derived the relation:
\begin{eqnarray}\displaystyle\label{CatZ}
{\rm M_{K} 
 =  -2.353\log P + 0.175\log Z - 0.597}
\end{eqnarray}
By using Eqs. 9 and 10 in \citet{Cat2004} and assuming [$\alpha$/Fe]$\sim$ 0.3 (e.g., \citealt{Carn1996}) we transformed Eq.~\ref{CatZ} into the form, presented in the second row of Table~\ref{ZP}. 

\citet{DO2004} derived an empirical relation between apparent $K$ magnitude and period for 21 RRab and 9 RRc stars in the LMC globular cluster Reticulum. \citet{DelP2006} obtained the relation between apparent $K_{\rm s}$ magnitude, metallicity and period  from the analysis of RR Lyrae stars in the Galactic globular cluster $\omega$ Cen. 

\citet{Sol2006} derived a ${\rm PL_{K}Z}$  relation by analysing 538 RR Lyrae stars in 15 Galactic clusters and in the LMC globular cluster Reticulum. This relation spans the metallicity range $-2.15<{\rm [Fe/H]}<-0.9$ dex. Mean $K$ magnitudes were estimated by combining Two-Micron-All-Sky-Survey (2MASS, \citealt{Cutri2003}) photometry and literature data. The zero-point was calibrated on RR Lyrae itself, whose distance modulus was derived using the {\it HST} trigonometric parallax measured for this star by \citet{Ben2002}.
\citet{Sol2008} presented $JKH$ time-series photometry of RR Lyrae and derived a new zero-point of \citet{Sol2006}'s ${\rm PL_{K}Z}$ relation.  

\citet{Bor2009} presented near-infrared $K_{\rm s}$ photometry and spectroscopically measured metallicity for a sample of 50 field RR Lyrae stars in inner regions of the LMC.  These authors  had 5 measurements in the $K_{\rm s}$ passband for most of the stars in their sample and used templates from \citet{Jon1996} to fit the light curves and derive the mean $K_{\rm s}$ magnitudes.   To improve statistics they added to their sample LMC RR Lyrae stars from \citet{Szew2008} dataset, and derived the ${\rm PL_{K_{\rm s}}Z}$ relation  based on the total sample of  107 LMC variables. The zero-point was calculated using \citet{Sol2008}'s mean $K$ magnitude, the  reddening and the  trigonometric parallax of RR Lyrae.

 \citet{Ben2011} recalibrated all the literature relations listed in Table~\ref{ZP}, but \citet{Cat2004}'s one, by fitting to  equations in the form:  $a({\rm logP} + 0.28) + b({\rm [Fe/H]} + 1.58) +ZP$,  the Lutz-Kelker-Hanson-corrected absolute magnitudes of the five MW RR Lyrae stars for which {\it HST} parallaxes are available. Since there are concerns about \citet{Ben2011} RR Lyrae star (namely RZ Cep) we transformed the ${\rm PL_{K}Z}$ relations of the LMC and MW RR Lyrae stars derived in this paper 
(Equations~\ref{M_LMC}, \ref{M_Ben_RZout}) to the form adopted by \citet{Ben2011} and determined their zero-points on the basis of \citet{Ben2011}'s  {\it HST} parallaxes but excluding RZ Cep.
The zero-points based on \citet{Ben2011} parallaxes are presented in column~5 of Table~\ref{ZP}, whereas the zero-point of our LMC  ${\rm PL_{K_s}Z}$ relation calculated by assuming the distance to the LMC (Subsection~\ref{zero-p}) and the zero-point of our MW ${\rm PL_{K}Z}$ relation based on the B-W studies are (Section~\ref{sec:MW}) are presented in Column~4. 

The slope in period of the RR Lyrae ${\rm PL_{K}Z}$ relations (Column~2 of Table~\ref{ZP})  differs significantly in the various studies. The slope we derived for the LMC RR Lyrae stars   is in excellent agreement with that derived by \citet{DelP2006}, whereas the slope of the MW RR Lyrae ${\rm PL_{K}Z}$ is in good agreement with that derived by \citet{Sol2006, Sol2008}.

The dependence on metallicity of the ${\rm PL_{K}Z}$ relations (Column~3 of Table~\ref{ZP}) also varies among different studies and generally is larger in the theoretical and semi-theoretical relations. 
The comparison of the metallicity dependence in the different empirical relations is complicated by the inhomogeneity of the  metallicity scales  adopted in these studies. 
Metallicities in \citet{DelP2006} are in the \citet{ZW1984} scale, whereas  in  \citet{Sol2006, Sol2008} are in the  \citet{CG1997} scale.
In the current study for the LMC RR Lyrae stars we used the metallicity scale defined by \citet{Grat2004} which is also the scale adopted by \citet{Bor2009}. As discussed in \citet{Grat2004} 
this scale is systematically 0.06 dex higher than Zinn \& West scale. This difference is small and systematic, hence should not affect the results of this comparison.
Finally, 
 for the MW RR Lyrae stars we used the metallicities measured by \citet{Clem1995}.
 Because the spectroscopic [Fe/H] values in \citet{Clem1995}  
 are derived from high dispersion spectra analyzed using standard reduction procedures, the derived metallicities are on the scale of the high dispersion spectra  (i.e., the \citealt{Carretta09} scale) and could be transformed to Zinn \& West  scale 
using the relations provided in \citet{Carretta09}.

The slope in metallicity of the ${\rm PL_{K}Z}$ relation based on the LMC RR Lyrae stars is the smallest among the various studies listed in Table~\ref{ZP} and it is close to \citet{Bor2009}'s slope. 
This is  consistent with the two studies both involving LMC variables and using exactly the same metallicity scale. 
The slope in metallicity we found for the MW RR Lyrae stars is larger than that of the LMC RR Lyrae stars and, in spite of the difference in metallicity scales,  it is very close  to the slope obtained by \citet{Sol2006} for  RR Lyrae stars in globular clusters. 
However, taken at face value, the metallicity slopes of the empirical relations in Table~\ref{ZP} appear to be all rather small  and in agreement to each other within the relative uncertainties, thus generally suggesting a mild dependence  the RR Lyrae ${\rm PL_{K}Z}$ independently on the specific environment.

 
\begin{center}
\begin{deluxetable}{lcccc}
\tabletypesize{\small}
\tablecaption{${\rm PL_{K_{\rm s}}Z}$ relations from the literature\label{ZP}}
\tablewidth{515pt}
\tablehead{
\colhead{Relation}&  \colhead{$a$}             &  \colhead{$b$}               & \colhead{ZP\tablenotemark{1} from the original relation}   & \colhead{ZP\tablenotemark{2} from \citet{Ben2011}}
}
\startdata
\cutinhead{{\it Theoretical or semi-theoretical relations}}
\citet{Bono2003} & -2.101 & $0.231 \pm 0.012$  & $-0.770 \pm 0.044$ & $-0.58 \pm 0.04$ \\
\citet{Cat2004} & -2.353  & 0.175  & -0.869 & - \\
\cutinhead{{\it Empirical relations}}
\citet{DO2004} &  $-2.16 \pm 0.09$    & - & - & $-0.56 \pm 0.02$  \\
\citet{DelP2006}\tablenotemark{3,}\tablenotemark{4}  & $-2.71 \pm 0.12$ & $0.12 \pm 0.04$ & - & $-0.57 \pm 0.02$ \\
\citet{Sol2006}\tablenotemark{5}        & $-2.38 \pm 0.04$ & $0.08 \pm 0.11$ & $-1.05 \pm 0.13$ & $-0.57 \pm 0.03$\tablenotemark{6} \\
\citet{Sol2008}\tablenotemark{5}        & $-2.38 \pm 0.04$ & $0.08 \pm 0.11$ & $-1.07 \pm 0.11$ & $-0.56 \pm 0.02$\\
\citet{Bor2009}\tablenotemark{3,}\tablenotemark{7} & $-2.11 \pm 0.17$ & $0.05 \pm 0.07$ & -1.05 & $-0.56 \pm 0.03$\\
This paper (LMC)\tablenotemark{3,}\tablenotemark{7} & $-2.73 \pm 0.25$ & $0.03 \pm 0.07$ & $-1.06 \pm 0.01$ & $-0.55 \pm 0.06$\tablenotemark{9}\\
This paper (MW)\tablenotemark{8} & $-2.53 \pm 0.36$ & $0.07 \pm 0.04$ & $-0.95 \pm 0.14$\ & $-0.56 \pm 0.06$\tablenotemark{9}\\
\enddata
\tablenotetext{1}{Zero-point of the original relation from the literature in the form: $a{\rm logP} + b{\rm [Fe/H]} +ZP$}
\tablenotetext{2}{Zero-point of the relation in the form: $a({\rm logP} + 0.28) + b({\rm [Fe/H]} + 1.58) +ZP$, as recalibrated  by \citet{Ben2011} }
\tablenotetext{3}{Near-infrared photometry in the  $K_{\rm s}$ band.}
\tablenotetext{4}{Metallicity is on the \citet{ZW1984} metallicity scale}
\tablenotetext{5}{Metallicity is on the \citet{CG1997} metallicity scale}
\tablenotetext{6}{Zero-point was calibrated by \citet{Ben2011} neglecting the metallicity term}
\tablenotetext{7}{Metallicity  on the scale  adopted by  \citet{Grat2004}, which is , on average, 0.06 dex higher than 
 \citet{ZW1984} scale}
\tablenotetext{8}{Metallicities from \citet{Clem1995},   
they are on the scale of the high dispersion spectra  (i.e., the \citealt{Carretta09})}
\tablenotetext{9}{ Zero-point was calibrated by us considering only four RR Lyrae stars (XZ Cyg, UV Oct, SU Dra and RR Lyr) from \citet{Ben2011} and excluding RZ Cep, since there are concerns about the parallax of this star}
\end{deluxetable}
\end{center}

\section{Gaia observation of RR Lyrae stars in the Milky Way\label{sec:gaia}}

The Gaia astrometric satellite will revolutionise many fields of astronomy \citep{gaia}. Of particular importance will be its catalogue of trigonometric parallaxes for more than one billion stars, with astrometric precision down to $\mu$as level. Due to Gaia's constant observation of the sky over the five year nominal mission, Gaia will repeatedly observe all stars brighter than its limiting magnitude, with an average of 70 observations per star. This will also make it possible for Gaia to discover and characterise many types of variables, including RR Lyrae stars  and Cepheids.

Gaia is observing in the broad visual band $G$ \citep{photometry} for its astrometric measurements, and is therefore not ideal for characterising the RR Lyrae ${\rm PLZ}$ relation, which exists only in the infrared passbands. However, since Gaia will provide accurate parallaxes for an expected tens of thousands of  MW RR Lyrae stars, it could serve as a perfect tool for the determination of the zero-point of the ${\rm PL_{K_{\rm s}}Z}$ relation through a combination with external datasets.
As it was discussed in Subsection~\ref{sec:ZP_Ben}, the current largest limiting factor in zero-point calibration of ${\rm PL_{K_{\rm s}}Z}$  and ${\rm M_{V}-{\rm [Fe/H]}}$ relations is the lack of a reliable and statistically significant sample of parallax measurements. The current state of the art is the sample of five RR Lyrae parallaxes from \cite{Ben2011} using the {\it HST}. Gaia will improve this situation by several orders of magnitude in both precision and numbers of objects.   Moreover, the distance to the LMC will be determined through the combination of Gaia parallaxes for a large sample of LMC bright stars, hence, a zero-point of the ${\rm PL_{K_{\rm s}}Z}$ relation based on the distance to the LMC will be obtained with a high precision.

\subsection{Method}
\label{MLE}
Using parallax data for calibration of a PL relation is complicated by the presence of statistical biases (e.g. \citealt{LK1973}) and sample selection effects (e.g. \citealt{Malmquist}). Non-linear transformations on the parallax cause a highly asymmetric uncertainty on the absolute magnitude when calculated using parallax and apparent magnitude information via the relation: 
m $-$ M =$-5 -  5 \log{\rm (\varpi)}$, where $\varpi$ is the parallax. 
Additionally, stars with a negative parallax measurement can not be used to calculate an absolute magnitude, though they do contain information. For these reasons, calculating an absolute magnitude for each star and fitting a PL relation directly to period and absolute magnitude leads to a biased result.

An unbiased solution can be achieved through modelling the data and inferring the slope and zero-point of the relation via statistical methods. For a catalogue of $N$ stars we can define $\bm{x}=(\bm{x}_1, \bm{x}_2, ..., \bm{x}_N)$ where the vector $(\bm{x}_i=m,l,b,P,\varpi,A)$ describes the observed data on each object. $P$ is the period, $m$ the apparent magnitude, $l$ and $b$ the position, $\varpi$ the parallax, and $A$ the extinction). We can additionally define that the vector $(\bm{x}_0=m_0,l_0,b_0,P_0,r_0,A_0)$ gives the `true' underlying object properties. 

We assume that the stars follow a PL relation of the form $M_0=\rho \textrm{log}P_0 +\delta$, although this can be changed to include other terms, such as metallicity, as needed. We can therefore model the true absolute magnitudes of the population as being Normally distributed around the PL relation, with the dispersion describing the intrinsic scatter on the relation:
\begin{equation} 
\varphi_{M}  (\bm{x}_0|\rho,\delta,\sigma_{PL}) = \frac{1}{\sigma_{PL}\sqrt{2\pi}}   e^{-0.5 \left( \frac{M_0- ( \rho \textrm{log}P_0 +\delta)}{\sigma_{PL} }\right)^2}
\label{eqn:Mmodel}
\end{equation}
where $\sigma_{PL}$ is the intrinsic dispersion of the PL relation. The parameters $\rho$ and $\delta$ are the slope and zero-point of the PL relation, which are to be found.

The true absolute magnitude is calculated through:
\begin{equation} 
M_0 = m_0 + 5\log(\varpi_0)+5  - A_0
\end{equation}

The observations are Normally distributed around the true values with a standard deviation given by the formal error on the measurement:
\begin{equation} 
\mathcal{E} (\bm{x}|\bm{x}_0) = \frac{1}{\epsilon_\varpi \epsilon_m \epsilon_A  \left(2\pi\right)^{3/2} } e^{-0.5 \left( \frac{\varpi-\varpi_0}{\epsilon_\varpi}\right) ^2 } e^{-0.5 \left( \frac{m-m_0}{\epsilon_m}\right) ^2 }  e^{-0.5 \left( \frac{A-A_0}{\epsilon_A}\right) ^2 }   \delta(l_0,b_0,P_0) 
\end{equation}
Assuming negligible errors on the position and period, the observations are described by a delta function. The terms $\epsilon_\varpi$, $\epsilon_m$, $\epsilon_A$ are the formal errors on the parallax, magnitude and extinction.

With the above models defined, the joint probability density function for the observations is:
\begin{equation} 
\mathcal{P}(\bm{x}_i|\rho,\delta,\sigma_{PL}) = \mathcal{S}(\bm{x}) \int_{\forall \bm{x}_0}  \varphi_{M} (\bm{x}_0|\rho,\delta,\sigma_{PL})\mathcal{E} (\bm{x}|\bm{x}_0) d\bm{x_0}
\label{eqn:Likelihood}
\end{equation}
the `true' parameters $\bm{x}_0$ are never known and so these values are marginalised through integration. The term $\mathcal{S}(\bm{x})$ is the selection function, which takes the probability of observing a star into account, given the properties of the star and the instrument's observational capabilities. To take the fact that Gaia is a magnitude limited sample into account, a step function is used with 
\begin{equation} 
\mathcal{S}(\bm{x}) =\begin{cases}
    1, & \textrm{if $G$\textless 20}.\\
    0, & \textrm{otherwise}.
  \end{cases} 
  \end{equation}

The Maximum Likelihood Estimation of the parameters are found by maximising equation \ref{eqn:Likelihood} by varying the parameters ($\alpha$, $\rho$, $\sigma_{PL}$). This formulation avoids non-linear transformations on error effected data, and includes a selection function which avoids the Malmquist bias.

\subsection{Simulated Gaia data\label{sgd}}

In order to check the application of the method defined in Section~\ref{MLE} we have used the sample of 23 RR Lyrae stars in the MW discussed in Section~\ref{sec:MW}  (see also Table~\ref{tab:MW}). 
In order to investigate the performance of the Gaia satellite and the contents of the end-of-mission catalogue, Gaia's Data Processing and Analysis Consortium (DPAC) has a group working on the simulation of several aspects of the Gaia mission. One major product of this work is the Gaia Object Generator (GOG; \citealt{GOG}), designed to simulate both individual Gaia observations and the full contents of the end-of-mission catalogue.  GOG includes a full mathematical description of the nominal performance of the Gaia satellite, and is therefore capable of determining the expected precision in astrometric, photometric and spectroscopic observations. In general, the precision depends on the apparent magnitude of the star, its colour, and its sky position, which affects the number and type of observations made (due to the Gaia scanning law). 

To obtain a distance for each RR Lyrae star from the sample, we use:
{
\begin{equation}
\label{plmforgog}
M_{K} = -2.53~ \textrm{log}P  - 0.95
\end{equation} 
as determined 
in Equation~\ref{M_MW} to obtain an absolute magnitude (neglecting the metallicity term for simplicity). We then determine a distance by combining this absolute magnitude with the apparent magnitude and extinction as defined above. Colour information as (V-I) is obtained from the Hipparcos catalogue \citep{hipparcos} where available. The apparent magnitude, position, colour, and period data form the basis of a synthetic catalogue of RR Lyrae stars, along with the distance obtained from the 
${\rm PM_{K}Z}$ 
 relation, and is used as the input catalogue of `true' parameters for GOG. 
 
GOG then creates simulated Gaia observations for our sample. We take the ${\rm PM_{K}Z}$  
elation (Equation~\ref{plmforgog}) as true, as a study of the possible precision in 
${\rm PM_{K}Z}$  
 calibration after the Gaia data will become available.

Using the fitting method described in Section~\ref{MLE} to the data including the simulated parallax observations and simulated errors applied to parallax and apparent magnitude, we find a ${\rm PM_{K}}$ relation of:

\begin{eqnarray}\displaystyle\label{templ}
M_{K} =  (-2.531 \pm 0.038)\log P + (-0.95\pm 0.01) 
\end{eqnarray}

 Comparison of these results to the input relation shows very good agreement. It proves that the fitting procedure given in Section~ \ref{MLE} is accurate and unbiased.
 When Gaia parallaxes will become available for the much larger sample of RR Lyrae stars, we will apply the described method to fit the  ${\rm PL_KZ}$ relation of RR Lyrae variables in the MW. Moreover, precise distance to the LMC obtained from the combination of Gaia parallaxes for a large sample of the bright LMC stars, will allow us to calibrate zero-point of the ${\rm PL_{K_{\rm s}}Z}$ relation based on the LMC RR Lyrae stars.
This provides a flavor of what will be possible to achieve with Gaia parallaxes.

\section{Summary}

We studied  a sample of 70 RR Lyrae stars in the LMC, for which multi-epoch  $K_{\rm s}$ photometry from the VMC survey, precise periods from the OGLE~III catalogue and spectroscopically determined metallicities \citep{Grat2004}, are available. There are 13 epoch data in the  $K_{\rm s}$ band for all stars in the sample, that allowed us to determine mean  $K_{\rm s}$ magnitudes with a great accuracy. 

Specifically for this work we developed a fitting approach. This method has several advantages compared to the Minimum-Least-Squares fitting, such as taking into account potentially significant intrinsic dispersion of the data, non-negligible errors in two dimensions and the possibility of inaccuracies in the formal error estimates. We used this method to derive the ${\rm PL_{K_{\rm s}}Z}$ relation of the 70 RR Lyrae stars in the LMC. Potentially the method could be used to fit any other sample of data.

The zero-point of the derived ${\rm PL_{K_{\rm s}}Z}$ relation was estimated in two different ways: (i) by assuming the distance to the LMC determined 
  by \citet{Pietr2013}; (ii) by applying {\it HST} parallaxes of  four MW RR Lyrae stars by  \citet{Ben2011}. The zero-point derived using the  MW RR Lyrae stars is 0.2 mag larger and, consequently, gives a longer distance to the LMC: $(m-M)=18.68\pm0.10$ mag. In future studies we suggest to use the relation based on the precise distance to the LMC:

\begin{eqnarray}
\label{sum_plmforgog}
M_{K_{\rm s}} 
& = &  (-2.73 \pm 0.25) \textrm{log}P + (0.03 \pm 0.07){\rm [Fe/H]}\nonumber\\
& - & (1.06 \pm 0.01)
\end{eqnarray} 

We found a negligible dependence of the $M_{K_{\rm s}}$ on metallicity, which could be caused by the relatively small range in metallicity covered by the LMC RR Lyrae stars. Thus, we applied the fitting approach to 23 RR Lyrae stars in the MW, for which absolute $M_K$ and $M_V$ magnitudes are known from Baade-Wesselink studies. We derived the ${\rm PL_{K_{\rm s}}Z}$ relation for MW RR Lyrae stars in the form:  

\begin{eqnarray}\displaystyle \label{sum:M_MW}
M_{K} 
& = & (-2.53 \pm 0.36) \textrm{logP} + (0.07 \pm 0.04){\rm [Fe/H]}\nonumber\\
& - & (0.95\pm 0.14)
\end{eqnarray}

Even though the metallicities of the MW RR Lyrae stars span a wide range [-2.5; 0.17] dex, the dependence on metallicity is relatively small and consistent, within the errors,  with the slope in metallicity found for the LMC RR Lyrae variables. We concluded that the small range of metallicities doesn't cause the negligible dependence of the $M_K$ on metallicity for the LMC RR Lyrae stars.  


To solve the problem of the ${\rm PL_{K_{\rm s}}Z}$ zero-point, a large sample of RR Lyrae stars with precisely determined parallaxes is necessary. A great contribution to this field is expected by the  Gaia satellite. By using GOG we simulated Gaia parallaxes of 23 MW RR Lyrae stars with observational errors. We present  a method for the calibration of the PL relation which avoids several of the problems which arise when using parallax data. The method was tested by deriving the ${\rm PL_{K_{\rm s}}}$ relation based on the simulated Gaia parallaxes. When combined with metallicity and photometry from other sources and a statistical tool such as the one developed in the present study, the extraordinary large sample of Gaia parallaxes for RR Lyrae stars will allow us to estimate these relations with unprecedented precision.

\acknowledgments
\section*{Acknowledgements}
This work was supported by the Gaia Research for
European Astronomy Training (GREAT-ITN) Marie Curie network, funded
through the European Union Seventh Framework Programme [FP7/2007-2013]
under grant agreement n. 264895. This work was also supported by the MINECO (Spanish Ministry of Economy) - FEDER through grant ESP2013-48318-C2-1-R. We thank the Cambridge Astronomy Survey Unit (CASU) and the Wide Field Astronomy Unit
(WFAU) in Edinburgh for providing calibrated data products under the support of the
Science and Technology Facility Council (STFC) in the UK.
We are grateful to P. Montegriffo for the development and maintenance of the GRATIS software.
We thank the members of the OGLE team for making public their catalogues.

\appendix

\section{Bayesian fitting approach}
\label{derivation}

This method is based on the prescription of \cite{Hogg2010}, extended into three dimensions and implemented in Python using a Markov Chain Monte Carlo sampler to obtain parameter estimates along with their complete posterior PDF. Initially, we model the data as being drawn from a thin plane defined by:
\begin{equation}
f(x,y) = A\,x + B\,y + C 
\end{equation}
where $A$ is the slope in the $x$ axis, $B$ is the slope in the $y$ axis, and $C$ is the intercept. In this initial model we assume that we have data in three axis, $x$, $y$ and $z$, with errors only in the $z$ axis.

In this model, given an independent position ($x_i$,$y_i$), an uncertainty $\sigma_{zi}$, slopes $A$ and $B$, and an intercept $C$, the frequency distribution $p(z_i|x_i,y_i,\sigma_{zi},A,B,C)$ for $z_i$ is
\begin{equation}\label{eq:objectivei}
p(z_i|x_i,y_i,\sigma_{zi},A,B,C) = \frac{1}{\sqrt{2\,\pi\,\sigma_{zi}^2}}
 \,\exp\left(-\frac{[z_i - A\,x_i -B\,y_i - C]^2}{2\,\sigma_{zi}^2}\right) 
\end{equation}
Therefore the likelihood is defined as:
\begin{equation}\label{eq:like}
\like = \prod_{i=1}^N \ p(z_i|x_i,y_i,\sigma_{zi},A,B,C) 
\end{equation}
Taking the logarithm,
\begin{equation}
\ln\like =  K - \sum_{i=1}^N \frac{[z_i - A\,x_i -B\,y_i - C]^2}{2\,\sigma_{zi}^2} 
\end{equation}
which is effectively the least-squares solution. K is a normalisation coefficient. Returning to Bayes rule it is possible to define:
\begin{equation}
p(A,B,C|\allz,I) = \frac{p(\allz|A,B,C,I)\,p(A,B,C|I)}{p(\allz|I)} 
\end{equation}
$\allz$ is all the data $z_i$. $I$ is all of the information of $x$ and $y$, $\allxy$, along with the formal errors $\allerrors$, plus any other prior information which may be available.  In our case we use uninformative (uniform) priors, making our inference method analogous to Maximum Likelihood Estimation.

\subsection{Multiple errors, no dispersion}
  
As in this case there exist errors in more than one axis, they can be put together into a covariance tensor $\mS_i$
\begin{equation}
\mS_i \equiv \left[\begin{array}{ccc}
\sigma_{xi}^2 & \sigma_{xyi} & \sigma_{xzi} \\ 
\sigma_{xyi} & \sigma_{yi}^2 & \sigma_{yzi} \\
\sigma_{xzi} & \sigma_{yzi} & \sigma_{zi}^2 \\
\end{array}\right] 
\end{equation}
With errors in several dimensions, our observed data point ($x_i$,$y_i$,$z_i$) could have been drawn from any true point along the plane ($x$,$y$,$z$). Making the probability of the data, given the model and the true position:
\begin{equation}
p(x_i,y_i,z_i|\mS_i,x,y,z) = \frac{1}{2\,\pi\,\sqrt{\det(\mS_i)}}
  \,\exp\left(-\frac{1}{2}\,\transpose{\left[\mZ_i - \mZ\right]}
  \,\inverse{\mS_i}\,\left[\mZ_i - \mZ\right]\right) 
\end{equation}
where we have implicitly made column vectors
\begin{equation}\label{eq:mZ}
\mZ = \left[\begin{array}{c} x \\ y \\ z\end{array}\right] \quad ; \quad
\mZ_i = \left[\begin{array}{c} x_i \\ y_i \\ z_i\end{array}\right] .
\end{equation}
In only two dimensions (e.g. $y$ and $z$), the slope (e.g. $B$) can be described by a unit vector $\vhat$ \emph{orthogonal} to the line or linear relation (at any $x$):
\begin{equation}
\vhat
 = \frac{1}{\sqrt{1+B^2}}\,\left[\begin{array}{c}-B\\1\end{array}\right]
 = \left[\begin{array}{c}-\sin\theta\\\cos\theta\end{array}\right] 
\end{equation}
where the angle $\theta = \arctan B$ is made between the line and the $y$ axis. The orthogonal displacement $\Delta_i$ of each data point $(y_i,z_i)$ from the line is given by
\begin{equation}
\Delta_i = \transpose{\vhat}\,\left[\begin{array}{c}  y_i \\ z_i\end{array}\right] - C\,\cos\theta 
\end{equation}
Instead of extending fully into three dimensions, we will assume a negligible error in $x$ (which will be the period, so has justifiably higher precision). The value of $x$ can therefore be input directly into $\Delta_i$ without worrying about the interplay between the other parameters.
\begin{equation}
\Delta_i = \transpose{\vhat}\,\left[\begin{array}{c}  y_i \\ z_i\end{array}\right] -  ( C\,\cos\theta  +A \, x_i) 
\end{equation}
Assuming negligable errors in $x$ also redefines the covariance matrix of the errors as:
\begin{equation}
\mS_i \equiv \left[\begin{array}{ccc}
 \sigma_{yi}^2 & \sigma_{yzi} \\
\sigma_{yzi} & \sigma_{zi}^2 \\
\end{array}\right] 
\end{equation}
Similarly, each data point's covariance matrix $\mS_i$ projects down to an orthogonal variance $\Sigma_i^2$ given by
\begin{equation}\label{eq:Sigma}
\Sigma_i^2 = \transpose{\vhat}\,\mS_i\,\vhat
\end{equation}
and then the log likelihood for $(A,B,C)$ or $(A,\theta,C\,\cos\theta)$ can be written as
\begin{equation}\label{eq:twodlike}
\ln\like = K - \sum_{i=1}^N \frac{\Delta_i^2}{2\,\Sigma_{i}^2} 
\end{equation}
where $K$ is some constant. This likelihood can be maximized to find $A$, $B$ and $C$.

\subsection{Dispersion}

The final step is to introduce an intrinsic variance in the line, V, orthogonal to the line.

According to \cite{Hogg2010}, each data point can be treated as being drawn from a projected distribution function that is a convolution of the projected uncertainty Gaussian, of variance $\Sigma_i^2$ defined above, with the Gaussian intrinsic scatter of variance $V$.  Therefore the likelihood becomes:
\begin{equation}
\ln\like = K - \sum_{i=1}^N \frac{1}{2}\,\ln(\Sigma_{i}^2+V)
 - \sum_{i=1}^N \frac{\Delta_i^2}{2\,[\Sigma_{i}^2+V]} 
\end{equation}
where again $K$ is a constant, everything else is defined as above. We then solve for $A$, $B$, $C$, and $V$ by maximising the log likelihood. The optimisation is performed using the adaptive MCMC sampler EMCEE developed by \cite{emcee}.  Any optimisation algorithm (e.g. Nelder-Mead, Powell, etc.) will find the maximum of the log likelihood. MCMC was chosen due to the evaluation of the full posterior PDF of the parameters, which is useful for the determination of formal errors.

\section{Metallicities for the MW RR Lyrae stars analyzed with the B-W method}
In this Appendix we provide a summary of spectroscopic metal abundances ([Fe/H]) derived for the field RR Lyrae stars analyzed with the B-W
 technique, by studies mainly based on high resolution spectroscopic material. Exceptions are the  values with reference 1,  2 and 4 that 
 come from compilations which include metallicities measured from low resolution spectroscopic data and photometric indices (see discussion in 
 Section 4.2 of Cacciari et al. 1992). 
 
  By applying our fitting approach we derived the ${\rm PM_KZ}$ relation for 23 MW RR Lyrae stars described in Section~\ref{sec:MW} adopting as an alternative the metallicity values from \citet{Lambert1996}:
 
 \begin{eqnarray}\displaystyle \label{M_MW_Lamb}
M_{K} 
& = & (-2.66 \pm 0.36) \textrm{logP} + (0.05 \pm 0.04){\rm [Fe/H]}\nonumber\\
& - & (1.00 \pm 0.15)
\end{eqnarray}
 
The intrinsic dispersion of the relation is found to be  0.007 mag. The RMS deviation of the data around the relation, neglecting the intrinsic dispersion, is 0.090 mag. The slope in logP of the relation based on \citet{Clem1995}'s metallicities (Eq.~\ref{M_MW}) differs from the slope obtained using \citet{Lambert1996}'s metallicities (Eq.~\ref{M_MW_Lamb}), however the values are consistent within the errors.

\begin{deluxetable}{lccl}
\tabletypesize{\footnotesize}
\tablecaption{Literature metallicities of the MW RR Lyrae stars analyzed with the B-W method\label{tab:MWFe}}
\tablewidth{300pt}
\tablehead{
\colhead{Star} & \colhead{[Fe/H]} & \colhead{$\sigma_{\rm [Fe/H]}$}& Reference\\
{}&{(dex)~~}&{(dex)~~}&\\
}
\startdata

UU Cet    &   $-$1.38	&   0.08  &  (5), average of FeI and FeII\\
                 &   $-$1.33 &   0.08  & (3), average of FeI and FeII\\
                 &   $-$0.95 &   0.16  & (7*), re-calibration of $\Delta$S and $\Delta$s value adopted by (7)\\
                 &   $-$1.45 &   0.16  & (7*), re-calibration of $\Delta$S in (7) and $\Delta$s value adopted by (5)\\
                 &   $-$1.20 &   0.20 &   (4)\\  
                 &   $-$1.28 &            &  (1,2)\\  
                 &                  &            &           \\
SW And   &  $-$0.06	&   0.08  &  (5), average of FeI and FeII\\
                 &  $-$0.41  &  0.10  &   from FeI in (7)\\
                 &  $-$0.24  &  0.12  &   from FeII in (7)\\
                 &  $-$0.34	&   0.085  (s=0.12) & average of FeI and FeII in (7)\\
                 &  $-$0.27  &  0.15  &  (11),  from FeI and solar abundance 7.51\\
                 &  $-$0.24  &  0.15  &  (11),  from FeII and solar abundance 7.51\\
                 &  $-$0.255&  0.15 &  (11). average of FeI and FeII\\
                 &  +0.20      &  0.08 &  (3), average of FeI and FeII\\
                 & $-$0.16   &           &(9)\\
                 & $-$0.15   &0.15    & (4)\\
                 &  $-$0.24  &           &   (1,2)\\
                 &                  &            &           \\
RR Cet    &  $-$1.38	&  0.08  &  (5), average of FeI and FeII \\
                 &   $-$1.62 &  0.15  & (11), from FeI\\
                 &   $-$1.49 &  0.15  & (11), from FeII\\
                 &   $-$1.18 &  0.09  & (3), average of FeI and FeII\\
                 &  $-$1.61   &           &(9)\\
                 &  $-$1.36  & 0.16  & (7*), re-calibration of $\Delta$S and $\Delta$s value adopted by (7)\\ 
                 &  $-$1.25   & 0.10    & (4)\\
                 &  $-$1.45  &           &   (1,2)\\
                 &                  &            &           \\
X Ari	        &  $-$2.50	&   0.08  &  (5), average of FeI and FeII \\ 
                  &$-$2.19   &  0.17   &  (6),  from FeI\\                                                                    
                  &$-$2.54   & 0.09    &  from FeI in (7)\\                                                                     
                  &$-$2.75    &0.08    &  from FeII in (7)\\ 
                  &$-$2.66     &0.105 (s=0.15) & average of FeI and FeII in (7)\\ 
                  &$-$2.74    & 0.09  & (3), average of FeI and FeII \\ 
                  &$-$2.61     &           &    (8), from FeI\\
                  &$-$2.62     &           &    (8), from FeII\\
                  &$-$2.68     &           &     (9)\\
		&$-$2.20    &0.10    &(4)\\
                  &$-$2.43     &           &     (1,2)\\
                  &                  &            &           \\
AR Per     &$-$0.34	&  0.16  &  (5*), re-calibration of $\Delta$S and $\Delta$s value adopted in  (5)\\ 
                  &$-$0.23   &  0.07  &  from FeI in (7)\\                                                                    
                  &$-$0.41   &  0.08  &  from FeII in (7)\\                                                                     
                  &$-$0.31   &  0.09  (s=0.13)&  average of FeI and FeII in (7)\\ 
                  &$-$0.24  &  0.15  & (11) from FeI\\
                  &$-$0.29  &  0.15  & (11) from FeII\\
                  &$-$0.32   &           &     (9)\\
	         &$-$0.30   &  0.20    &(4)\\
                  &$-$0.30   &           &     (1,2)\\
                  &                  &            &           \\
RX Eri      &$-$1.63  & 0.16  &  (5*), re-calibration of $\Delta$S and $\Delta$S value adopted in (5)\\ 
                  &$-$1.98  & 0.16  &  (7*), re-calibration of $\Delta$S and $\Delta$S value adopted in (7)\\
	         &$-$1.40  &  0.20    &(4)\\
                  &$-$1.33  &           &     (1,2)\\
                 &                 &            &           \\
RR Gem  &$-$0.32	&  0.16  &  (5*),  re-calibration of $\Delta$S and $\Delta$S value adopted in (5)\\
                 &$-$0.44    & 0.16   & (7*),  re-calibration of $\Delta$S and $\Delta$S value adopted in (7)\\
	       &$-$0.30   &  0.25    &(4)\\
                &$-$0.29   &           &     (1,2)\\
                 &                  &            &           \\
TT Lyn     &$-$1.64	&   0.16  &  (5*), re-calibration of $\Delta$S and $\Delta$S value adopted in (5)\\ 
                  &$-$1.63   &  0.08   &  from FeI in (7)\\                                                                    
                  &$-$1.33   &  0.06   &  from FeII in (7)\\                                                                     
                  &$-$1.44     &0.150 (s=0.21) & average of FeI and FeII in (7)\\ 
                  &$-$1.64   &0.15     & (11),  from FeI\\
                  &$-$1.53   &0.15     & (11),  from FeII\\
                  &$-$1.41     &           &     (9)\\
 		&$-$1.35    &0.20    &(4)\\
                  &$-$1.56     &           &     (1,2)\\
                  &                  &            &           \\
T Sex       & $-$1.20	&   0.16  &  (5*), re-calibration of $\Delta$S and $\Delta$S in (5)\\
                  &$-$1.75   &  0.12   &  from FeI in (7)\\                                                                    
                  &$-$1.50   &  0.09   &  from FeII in (7)\\                                                                     
                  &$-$1.59     &0.125 (s=0.18) & average of FeI and FeII in (7)\\ 
		&$-$1.20   &0.15    &(4)\\
                  &$-$1.34     &           &     (1,2)\\
                  &                  &            &           \\
RR Leo  & $-$1.37	&   0.16  &  (5*), re-calibration of $\Delta$S and $\Delta$S value in (5)\\
                  &$-$1.54   &  0.11   &  from FeI in (7)\\                                                                    
                  &$-$1.17   &  0.10    &  from FeII in (7)\\                                                                     
                  &$-$1.34     &0.185 (s=0.26) & average of FeI and FeII in (7)\\ 
                  &$-$1.39     &           &     (9)\\
		&$-$1.15    &0.20   &(4)\\
                  &$-$1.60     &           &     (1,2)\\
                 &                  &            &           \\
WY Ant   & $-$1.32	&   0.16  &  (5*), re-calibration of $\Delta$S and $\Delta$S  value in (5)\\
                & $-$1.96   &   0.10 & (10), from FeI\\                      
                & $-$1.96    &   0.10 & (10), from FeII\\                     
                & $-$1.96    &   0.10 &  average of FeI and FeII in (10) + errors from us \\                     
                 & $-$1.36   &0.16   & (7*), re-calibration of $\Delta$S and $\Delta$S  value in (7)\\ 
                 &$-$1.25    & 0.20   &(4*)\\
                 &$-$1.48     &           &     (1,2)\\
                  &                  &            &           \\
W Crt	&$-$0.89	&   0.16  &  (5*), re-calibration of $\Delta$S and $\Delta$S value  in (5)\\
                   &$-$0.91  &   0.16  & (7*),  re-calibration of $\Delta$S and $\Delta$S  value in (7)\\
		&$-$0.70   &0.20    &(4*)\\
                  &$-$0.54     &           &     (1,2)\\
                 &                  &            &           \\
TU UMa  &$-$1.38	&   0.16  &  (5*), re-calibration of $\Delta$S and $\Delta$S  value in (5)\\
                  &$-$1.64   &  0.08   &  from FeI in (7)\\                                                                    
                  &$-$1.45   &  0.08    &  from FeII in (7)\\                                                                     
                  &$-$1.55     &0.095 (s=0.13) & average of FeI and FeII in (7)\\ 
		&$-$1.31    &0.05 or 0.14  &(6), from FeI, error of 0.05 likely a typo\\
                   &$-$1.72   &0.15  & (11),  from FeI\\
                   &$-$1.57   &0.15  & (11),  from FeII\\
                  &$-$1.46     &           &     (9)\\
		&$-$1.25    &0.20   &(4)\\
                  &$-$1.51     &           &     (1,2)\\
                  &                  &            &           \\
UU Vir  &  $-$0.64	&   0.16  &  (5*), re-calibration of $\Delta$S and $\Delta$S  value in (5)\\
                  &$-$0.85   &  0.12   &  from FeI in (7)\\                                                                    
                  &$-$0.79   &  0.07    &  from FeII in (7)\\                                                                     
                  &$-$0.81     &0.03 (s=0.04) & average of FeI and FeII in (7)\\ 
                  &$-$0.90     &           &     (9)\\
		&$-$0.50    &0.15   &(4)\\
                  &$-$0.87     &           &     (1,2)\\
                  &                  &            &           \\
SW Dra  &$-$0.91	&   0.16  &  (5*), re-calibration of $\Delta$S and $\Delta$S  value in (5)\\
                 &$-$1.37    &  0.15   & (11),  from FeI\\
                 &$-$1.28    &  0.15   & (11),  from FeII\\
                 &$-$0.81    &  0.16   & (7*), re-calibration of $\Delta$S and $\Delta$S  value in (7)\\
		&$-$1.14    &0.20   &(4)\\
                  &$-$1.12     &           &     (1,2)\\
                  &                  &            &           \\
RV Oct  & $-$1.92	&   0.16  &  (5*), re-calibration of $\Delta$S and $\Delta$S  value in (5)\\
                & $-$1.54    &   0.10  & from FeI in (10) errors from us\\                     
                & $-$1.54    &   0.11  & from FeII in (10) errors from us\\                     
                & $-$1.54    &   0.11  & average of FeI and FeII in (10) errors from us\\ 
                & $-$1.98    &   0.16  & (7*),   re-calibration of $\Delta$S in (7) and $\Delta$S  value in (5)\\ 
 	       &$-$1.75    & 0.20   &(4*)\\
                 &$-$1.71     &           &     (1,2)\\
                 &                  &            &           \\
TV Boo    &$-$2.31	&   0.16  &  (5*), re-calibration of $\Delta$S and $\Delta$S  value in (5)\\
                   &$-$2.36  &   0.16  &  (7*),  re-calibration of $\Delta$S in (7) and $\Delta$S  value in (5)\\
		&$-$2.30  &   0.15   &(4)\\
                  &$-$2.44     &           &     (1,2)\\
                  &                  &            &           \\
RS Boo  & $-$0.37	&   0.16  &  (5*), re-calibration of $\Delta$S and $\Delta$S  value in (5)\\
                  &$-$0.55  &  0.11   &  from FeI in (7)\\                                                                    
                  &$-$0.39   &  0.08   &  from FeII in (7)\\                                                                     
                  &$-$0.45   &0.08 (s=0.11) & average of FeI and FeII in (7)\\ 
                  &$-$0.48   &0.15  & (11),  from FeI\\
                  &$-$0.33   &0.15  & (11),  from FeII\\
		&$-$0.40  &   0.25   &(4)\\
                  &$-$0.36     &           &     (1,2)\\
                  &                  &            &           \\
VY Ser  & $-$1.71	&   0.08  &  (5), average of FeI and FeII\\
                  &$-$2.09  &  0.08   &  from FeI in (7)\\                                                                    
                  &$-$1.76   &  0.06   &  from FeII in (7)\\                                                                     
                  &$-$1.88   & 0.165 (s=0.23) & average of FeI and FeII in (7)\\ 
                  &$-$1.71   & 0.07    & (3),  average of FeI and FeII\\ 
                  &$-$2.00   & 0.15    & (11),  from FeI\\
                  &$-$1.90   & 0.15    & (11),  from FeII\\
		&$-$1.80  &   0.15   &(4)\\
                  &$-$1.79     &           &     (1,2)\\
                  &                  &            &           \\
V445 Oph&  +0.17	&   0.08  &  (5), average of FeI and FeII\\
                   & +0.13    &   0.10  & (3), average of FeI and FeII\\ 
                   &$-$0.26  &   0.16  &  (7*),  re-calibration of $\Delta$S in (7) and $\Delta$S  value in (7)\\
                  &$+$0.24     &           &     (9)\\
		&$-$0.30    &0.25   &(4)\\
                  &$-$0.19     &           &     (1,2)\\
                 &                  &            &           \\
TW Her  & $-$0.58	&   0.16  &  (5*), re-calibration of  $\Delta$S and $\Delta$S value in (5)\\
                &$-$0.56     &   0.16  &  (7*), re-calibration of  $\Delta$S and $\Delta$S value in (7)\\ 
	      &$-$0.50  &   0.15   &(4)\\
                &$-$0.69     &           &     (1,2)\\
                 &                  &            &           \\
AV Peg  &$-$0.03	&   0.16  &  (5*), re-calibration of  $\Delta$S and $\Delta$S value in (5)\\
                  &$-$0.35   &  0.09   &  from FeI in (7)\\                                                                    
                  &$-$0.04   &  0.06   &  from FeII in (7)\\                                                                     
                  &$-$0.14   &0.155  (s=0.22) & average of FeI and FeII in (7)\\ 
		&0.00        &   0.20   &(4)\\
                  &$-$0.08     &           &     (1,2)\\
                 &                  &            &           \\
RV Phe  & $-$1.69	&   0.16  &   (5*), re-calibration of  $\Delta$S and $\Delta$S value in (5)\\
                & $-$1.75    &   0.16  &   (7*), re-calibration of  $\Delta$S and $\Delta$S value in (7)\\ 
		&$-$1.35  &   0.25   &(4)\\
                  &$-$1.69     &           &     (1,2)\\
\enddata

\tablecomments{(1)  [Fe/H] from Table~1 in \citet{Fern1998a}; (2) [Fe/H]  values from  \citet{Feast2008};
 (3)   [Fe/H] values  from \citet{Nem2013} 
using the   values from the VWA analysis as they are listed in column (9) of Table~7 in that paper. They are the average of the FeI and FeII
     abundances.
 (4) [Fe/H] values from Table~11 of \citet{Cacc1992};  (4*)  [Fe/H] values from Table~16 of \citet{Ski1993};    
 (5) [Fe/H] values from abundance analysis of high resolution spectra performed by  \citet{Clem1995}. Values 
      are the average of the FeI and FeII measurements adopting for the solar abundance: log $\epsilon$(FeI)=7.56 
      and log $\epsilon$(FeII)=7.50, respectively  (see Table~12 of \citealt{Clem1995});  (5*)  [Fe/H] values obtained from \citet{Clem1995} re-calibration of 
      the $\Delta$S index.
%
(6) [Fe/H] values from Table~7 of \citet{Pancino2015}. Abundances are from FeI averaging values from different spectra of 
     the same star as described at the end of the paper Section 4.4. The error for the abundance of TU Uma is likely a typo, according to the paper 
     Table~6 likely should be 0.14 dex;
 (7) Metal abundances from \citet{Lambert1996}. 
      [Fe/H] values were derived from the photometric determinations in the paper Table~3 adopting for the sun log $\epsilon$(Fe)= 7.51 (according to what stated in the footnotes 
      of the paper Table~5); 
     (7*) [Fe/H] values obtained from \citet{Lambert1996}  re-calibration of 
      the $\Delta$S index (Equation 3 in that paper)  which was derived by these Authors using the FeII
     abundances and $\Delta$S from \citet{Blanco1992}.  \citet{Lambert1996} does not provide errors for the metallicities from $\Delta$S, hence we adopted an error of 
      0.16 dex,  as done by \citet{Clem1995}  for their metallicities from $\Delta$S;   
 (8)  [Fe/H] values from Table~ 10 of \citet{Haschke2012}; 
 (9) [Fe/H] values abundance analysis performed by  \citet{Wallerstein2010}, no errors are provided; 
(10)  [Fe/H] values  from \citet{For2011} 
obtained as the average weighted by errors of the values in the paper Table~5. Errors are the sum in quadrature 
of the individual errors in Table~5 divided by the square root of N (with N number of measurements, i.e.: 11 for WY Ant and 17 for RV Oct). The average value from FeI for
WY Ant published in Table~11 of \citet{For2011}. is $-$1.95 dex while we find $-$1.96 dex with our procedure. We also list the abundances from FeI and FeII separately, with errors 
     calculated as per  the above procedure; 
(11)  [Fe/H] values  from \citet{Fern1996}.
 They are from FeII with values taken from the paper Table~4b. Those from FeI are taken from the paper Table~4a. Errors are estimated
   by the authors to be of $\pm$0.15 dex in [Fe/H].}
 
  
\end{deluxetable}

\label{lastpage}

\end{document}